\documentclass[journal]{IEEEtran}

\ifCLASSINFOpdf
\else
\fi
\usepackage{amssymb}
\usepackage{amsmath}
\usepackage{cite}
\usepackage{xcolor}
\usepackage[ruled,vlined]{algorithm2e}
\usepackage{url}
\usepackage{graphicx}
\usepackage{gensymb}
\usepackage{caption}
\usepackage{subcaption}

\SetKwIF{If}{ElseIf}{Else}{if}{}{else if}{else}{end if}
\SetKwRepeat{Repeat}{do}{until}

\def\p{\mathbf{P}}
\def\d{\mathbf{D}}
\def\q{\mathbf{Q}}

\def\f{\mathbf{F}}
\def\x{\mathbf{x}}
\def\h{\mathbf{H}}

\begin{document}

\title{
 \textcolor{black}{
A Dual-Function Radar Communication System With  OFDM Waveforms and Subcarrier Sharing}
}
%
%
%

\author{Zhaoyi Xu,~\IEEEmembership{Student Member,~IEEE,}
        Athina Petropulu,~\IEEEmembership{Fellow,~IEEE,}
\thanks{This work was supported by NSF under grant ECCS-2033433}%
}


\maketitle

\begin{abstract}
A novel monostatic multiple-input multiple-output (MIMO) dual-function radar communication (DFRC) system is proposed, that uses the available bandwidth efficiently for both sensing and communication.
The proposed system, referred to as Shared-Subcarriers DFRC (SS-DFRC) transmits wideband, orthogonal frequency division multiplexing (OFDM) waveforms and allows the transmit antennas to use  subcarriers in a shared fashion.
A novel, low complexity target estimation approach is proposed to overcome the coupling of radar target parameters and transmitted symbols that arises in that case. When all subcarriers are used in a shared fashion, the proposed system achieves high communication rate but its sensing performance is limited by the size of the receive array. We show that when  some of the subcarriers are reserved for exclusive use by one  transmit antenna each (private subcarriers), the communication rate   can be traded off for improved sensing performance. The latter can be achieved  by using the private subcarriers to construct a larger aperture virtual array that yields higher resolution angle estimates.
The system is endowed with beamforming capability, via waveform precoding, where the precoding matrix is optimally designed to meet a joint sensing-communication system performance metric.

\end{abstract}


\vspace{-2mm}
\section{Introduction}

An emerging trend in  next-generation wireless applications \cite{6G_whitepaper} is to allow unconstrained access to spectrum for radar and communication systems for the purpose of increasing spectral efficiency. This has given rise to a lot of interest in designing systems that can coexist in the spectrum while using different platforms \cite{Babaei2013SpectrumSharing,Wang2015SpectrumSharing, Li2016SpectrumSharing}, or to  Dual Function Radar Communication (DFRC) systems that perform sensing and communication from a single platform \cite{Fan2020}. The former class can work with existing systems but requires means for controlling the interference between the two systems, for example, via a control center \cite{Li2016SpectrumSharing}. On the other hand, 
DFRC systems require  new signaling designs, but  do not require interference control between their radar and communication functionalities, and they further  offer reduced cost, lighter hardware, and lower power consumption. For those reasons, DFRC systems are of great interest to vehicular networks, WLAN indoor positioning,  unmanned aerial vehicle networks \cite{adas,jrc,5gvv,xu2020dfrc,Fan2020}. 
The contribution of this work falls along the lines of DFRC systems.
 
%

DFRC systems typically involve multi-antenna systems. There are various ways via which communication information can be embedded in the sensing process. It can be embedded in the way the waveforms are paired with transmit antennas \cite{sm,9093221,Wu2021FHDFRC,BaxterFHDFRC}, or in the  phase of the sidelobes in the array beampattern \cite{Euziere2014sidelobe}, or in the antenna activation pattern \cite{phase,9345999}. However, those approaches do not achieve high communication rate.

Communication information can also be directly embedded in the radar waveforms
\cite{mimodfrc1,ofdm,ofdmdfrc,xu2020dfrc,saddik2007UWdfrc,gaglione2016frftdfrc,hassanien2015phasedfrc,9345999,Liu2017OFDMDFRC,Dokhanchi2018OFDMDFRC,Wu2021FHDFRC}, and to enable high communication rate,
multi-carrier waveforms  have been considered in DFRC systems.
\textcolor{black}{Multi-carrier waveforms with Frequency-Hopping (FH)} were proposed in \cite{BaxterFHDFRC,Wu2021FHDFRC}, where the  bandwidth is divided into subbands, and each transmit antenna is paired with a subband in an exclusive fashion, with the pairing changing over time.
{However, this type of subbands assignment  uses only part of the available bandwidth, and thus yields lower resolution  target range information.}
Orthogonal frequency division multiplexing (OFDM) waveforms for DFRC systems have been explored in \cite{ofdm,ofdmdfrc,Liu2017OFDMDFRC,liu2019robust,liu2020super,Dokhanchi2018OFDMDFRC}.  
For communications, OFDM is a popular approach to achieve high communication rate and  deal with frequency selective fading. For those reasons, it  has been widely used in  wireless local area network (WLAN)\cite{Nee2006ofdm802.11}, power line communication (PLC)\cite{Li2020PLCOFDM} and 4G/5G mobile communications\cite{Ergen2009Mobilebroadband}. 
%
%

OFDM waveforms have also been used in radar, due to their 
ability to flexibly occupy the available spectral
resources \cite{ofdm}, and  easily overcome frequency selective propagation effects.
In \cite{ofdm,ofdmdfrc,Liu2017OFDMDFRC,liu2019robust,liu2020super}, the OFDM
subcarriers are assigned to antennas in an exclusive fashion. In that case,  the transmissions of different antennas are orthogonal, 
and a  virtual array can be constructed at the receiver  to obtain high  resolution angle estimates. \textcolor{black}{As compared to FH methods, OFDM methods  use all available bandwidth for sensing, which allows for  higher range resolution. However, due to the way carriers are assigned to antennas,  the communication bandwidth is not used efficiently.}

In this paper, we  propose a novel OFDM-based DFRC system that uses the available bandwidth efficiently for both sensing and communication.
It comprises a monostatic MIMO radar that transmits  precoded OFDM waveforms, allowing shared use of subcarriers among the transmit antennas. We will refer to the proposed system as 
Shared-Subcarriers DFRC (SS-DFRC).
In contrast to  prior  OFDM DFRC systems \cite{ofdm,ofdmdfrc,Liu2017OFDMDFRC,liu2019robust,liu2020super,Dokhanchi2018OFDMDFRC},  where each subcarrier   carries the data symbol of one antenna only,  the shared-used of subcarriers by the transmit antennas can achieve high communication rate.  
The challenge that arises when  the antennas share subcarriers  is  loss of waveform orthogonality between antennas, and further,  coupling of radar target parameters and communication symbols.
\textcolor{black}{
 When each subcarrier carries the symbol of one antenna only, the transmitted symbol can be easily canceled from  the received symbol, allowing for easy subsequent estimation of  angle, range and Doppler parameters \cite{ofdm}. However, when each subcarrier carries the symbols of multiple antennas, the received signal on each subcarrier is a superposition of multiple differently modulated  data symbols.
\textcolor{black}{One could  extract the target parameters via  maximum likelihood estimation,  as done in MIMO radar \cite{Tabrikian2008performance}, {by searching over the entire parameter space}.}
However, here  we propose  a novel, suboptimal but computationally simpler  approach to estimate the target parameters. }

\begin{figure}
  \centering\includegraphics[width=0.4\textwidth]{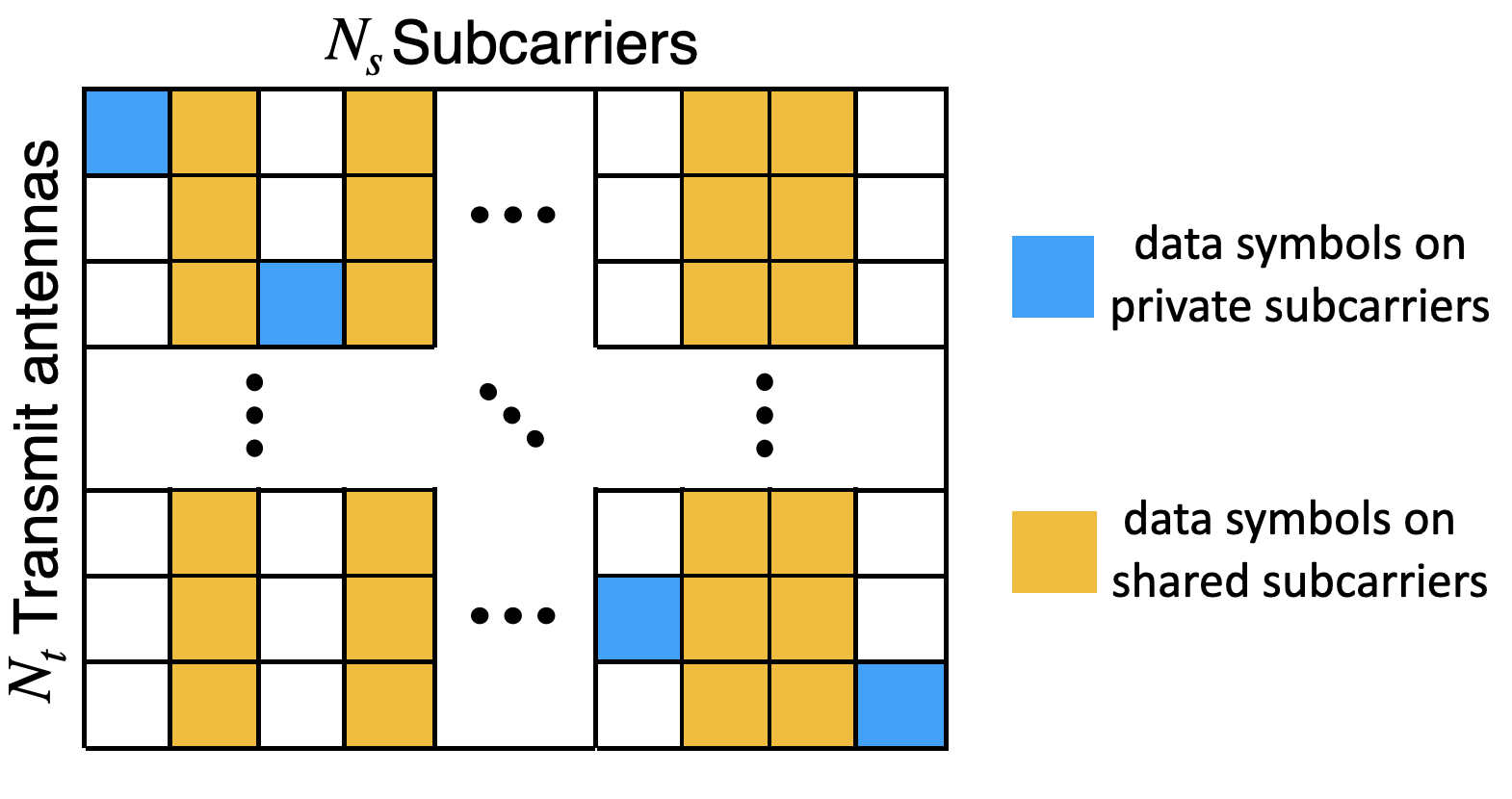}
  \caption{\footnotesize{Private and shared subcarriers}}\label{fig2}
  \vspace{-3mm}
\end{figure}

The proposed SS-DFRC system can operate with all subcarriers available to all transmit antennas, thus making the best use of the available bandwidth for communication. However, in that case,  the resolution  of  target angle estimation is limited by the aperture of the receive array. Further, 
angle errors affect the subsequent range and Doppler estimates. 
Also, due to waveform orthogonality loss,  the virtual array feature of MIMO radar cannot be exploited to improve angle resolution.
To address these shortcomings, we
 propose a way to trade off  communication rate for improved target estimation performance. In particular, we propose two types of subcarriers, namely  shared subcarriers, which can be used by all antennas simultaneously,  and  \textit{private} subcarriers, which are used by one antenna each (see Fig.~\ref{fig2}). The private subcarriers introduce  communication rate reduction. However,   
{the signal received on them 
can be effectively viewed as the response of a  virtual array (VA) that has higher aperture than the physical receive array. 
The VA steering vector depends on  target ranges as well as angles. Using the VA measurements, along with the range estimates obtained  on all (private and shared) subcarriers, and assuming that the target space corresponding to those range bins is sparsely populated by targets,  high resolution angle  estimates can be obtained  by formulating and solving a sparse signal  recovery (SSR) problem. The refined angles can then be used to improve the previously obtained range estimates.}


The proposed system is endowed with beamforming capability via waveform precoding.
The precoding  matrix is optimally designed to maximize a weighted combination of the radar beampattern performance and the communication signal-to-noise (SNR) ratio.
At the communication receiver side,
if the   receiver has more antennas that the  transmit array, and assuming channel knowledge, the transmitted data can be recovered via a least-squares method.


%

The novelty of the proposed work lies in the following:
\begin{enumerate}
    \item  
    A novel precoded OFDM DFRC system is proposed that fully utilizes the available bandwidth for communication by making all subcarriers available to all antennas. A  low complexity method   to estimate the target parameters is also proposed.
    \item A novel approach is proposed to
    flexibly trade off  communication rate for improved target estimation performance by introducing private as well as shared subcarriers.
    The signal on the private subcarriers can be used  construct a  virtual array with large aperture, thus enabling high target angle resolution.
    \item An iterative algorithm is proposed, which first estimates the target parameters based on shared and private subcarriers, and then uses the virtual array to refine the target parameters.
    \item The radar precoding matrix is optimally designed to meet a joint sensing-communication objective.

\end{enumerate}
Preliminary results of this work were reported in \cite{xu2020dfrc}. In addition to \cite{xu2020dfrc}, here, we (i) further exploit the frequency diversity for angle estimation on the receive array, (ii) explore the trade-off between sensing and communication, (iii) design an iterative scheme  based on the private subcarriers to refine radar target estimation, 
(iv) introduce precoding for beamforming purposes, and (v) propose a novel co-design  scheme that optimizes  a combination of  probing beampattern and communication SNR performances.

The remainder of this paper is organized as follows.
In Section II, we describe the target estimation process using precoded OFDM radar with all  subcarriers used as shared.
In Section III, we introduce the use of private subcarriers in addition to shared subcarriers,  discuss target parameter estimation using the virtual array, and present an iterative algorithm to refine target parameters. 
In section IV, we discuss the communication component of the proposed system. In Section V we formulate the precoder design problem by jointly optimizing sensing and communication performance. 
We provide simulation results on the system performance in Section VI and concluding remarks in Section VII.\\

\noindent{\textit {Notation}:} 
Throughout this paper, we use $\mathbb{R}$ and $\mathbb{C}$ to denote the sets of real and complex numbers, respectively.
$(\cdot)^T$ stands for the transposition operator, $(\cdot)^*$ denotes complex conjugate and $(\cdot)^H$ denotes complex conjugate transpose.
$||\cdot||_1$, $||\cdot||_2$, $||\cdot||_F$ represent  $\ell_1$, $\ell_2$ and Frobenius norms, respectively. ${\bf I}_N$ denotes an identity matrix of size $N\times N$.
 $\lfloor\cdot\rfloor$ denotes the floor function.

\IEEEpeerreviewmaketitle

\section{SS-DFRC with all subcarriers used as shared}

Let us consider a \textit{monostatic}, fully digital MIMO radar, comprising an $N_t$-antenna uniform linear array (ULA) to transmit, and an     $N_r$-antenna ULA to receive. The transmit and receive antennas spacing are
$g_t$ and $g_r$, respectively. 
The radar transmits precoded OFDM waveforms using $N_s$ subcarriers.
\textcolor{black}{In this section, we will assume that all subcarriers are available to  all transmit antennas, i.e., they are used  in a shared fashion.}

\begin{figure}
    \centering
    \vspace{-5mm}
    \includegraphics[width =7cm]{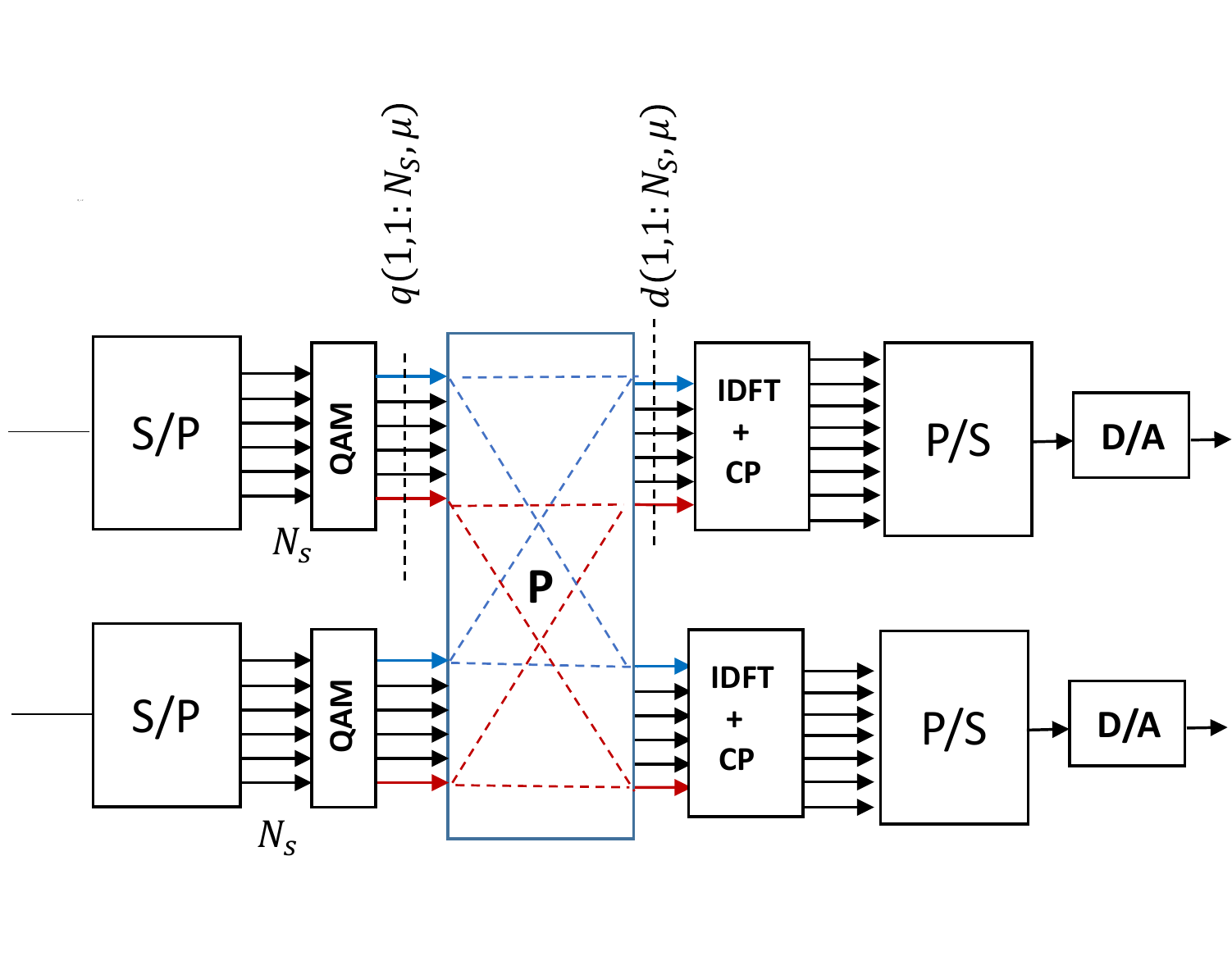}
    \vspace{-6mm}
    \caption{Beamforming with precoding on source data symbols}
    \label{fig:symbols}
    \vspace{-2mm}
\end{figure}

The radar transmitter is illustrated in Fig.~\ref{fig:symbols}.
The binary source data are divided into $N_t$ parallel streams,  and each stream is modulated 
and distributed to the OFDM subcarriers.
The outputs of all streams are  processed by a precoding matrix ${\bf P}$.
An inverse discrete Fourier transform (IDFT) is applied to   the data symbols of each stream,  a cyclic prefix (CP) is pre-appended to the result,  the samples are converted into an analog multicarrier signal with carrier frequency $f_c$ and are transmitted through the corresponding antenna. 
The latter signal  will be referred to as  OFDM symbol, and has duration $T_p$.

Let $\d\in\mathbb{C}^{N_t\times N_s}$ denote a matrix that contains the symbols to be transmitted during the $\mu$-th OFDM symbol, i.e.,
\begin{equation}
    {\d} = 
    \begin{bmatrix}
    d(0,0,\mu)&\dots&d(0,N_s-1,\mu)\\
    d(1,0,\mu)&\dots&d(1,N_s-1,\mu)\\
    \dots&\dots&\dots\\
    d(N_t-1,0,\mu)&\dots&d(N_t-1,N_s-1,\mu)
    \end{bmatrix}
\end{equation}
where  $d(n,i,\mu)$ denotes the symbol  transmitted by the $n$-th antenna, on the $i$-th subcarrier, during  the $\mu$-th OFDM symbol. 
\textcolor{black}{The $i$-th column of $\d$, i.e., ${\bf d}_i$, contains the symbols transmitted by all antennas on subcarrier $i$, while the $j$-th row of $\d$ contains the symbols transmitted by the $j$-antenna on all subcarriers. 
Unless otherwise indicated, the use of the symbol matrix will refer to one OFDM symbol, thus, for notational simplicity, the dependence on $\mu$ is not shown in the notation $\d$. } {We should note that, in the works without subcarrier sharing \cite{ofdm,ofdmdfrc,liu2020super}, each column of $\d$ contains only one nonzero element, while here the entries of $\d$ are nonzero data symbols.}

Let $\p\in\mathbb{C}^{N_t \times N_t}$ denote  the precoding matrix and $\q\in\mathbb{C}^{N_t\times N_s}$ the matrix containing the   data symbols in each OFDM symbol before precoding.
Then,
\begin{equation}
    \d = \p\q.
    \label{eq:transmitted symbols}
\end{equation}

The precoding scheme endows the system with beamforming capability, and the precoding matrix, ${\bf P}$, will be optimally designed to meet a certain performance metric.

The complex envelope of the transmitted baseband signal on the $i$-th subcarrier due to the $n$-th antenna equals
\begin{align}
    x(n,i,t) = \sum_{\mu=0}^{N_p-1} d(n,i,\mu) e^{j2\pi i{\Delta f} t}rect(\frac{t-\mu T_p}{T_p}), 
    \label{eqform}
\end{align}
for $n =1,...,N_t$ and $i = 0,1,\dots,N_s-1$, where $rect(t/T_p)$ denotes a rectangular pulse of duration $T_p$ and ${\Delta f}$ is the subcarrier spacing.

\textcolor{black}{Suppose that there are $K$ point targets in the far field, each characterized by angle $\theta_k$, range $R_k$, and Doppler frequency $f_{d_k}$.} It holds that $f_{d_k} = {2v_k f_c}/{c}$, with $c$ denoting the speed of light, and $v_k$ the radial velocity of the $k$-th target. 
The baseband equivalent of the signal reflected back and received by the $m$-th receive antenna on the $i$-th subcarrier is
\begin{equation}
        y(m,i,t)  = \sum_{k=1}^{K} \sum_{n=1}^{N_t}  \beta_{k} x(n,i,t-\tau^i_{kmn})e^{j2\pi f_{d_k}t}+ u_i(m,t),
        \label{receivedsig}
\end{equation}
for $m = 0,...,N_r-1$,  where \textcolor{black}{$ \beta_{k}$}
is a complex coefficient  accounting for the scattering process associated with the $k$-th target,  and
\begin{align*}
    \tau^i_{kmn}= 2R_k/c+(ng_t+mg_r)sin\theta_k/\lambda_i, 
\end{align*}
is the roundtrip delay of the $k$-th target, 
{with $\lambda_i$ the wavelength of the $i$-th subcarrier},
and $u_i(m,t)$ denoting noise or clutter.

Eq. \eqref{receivedsig} represents the effect of a single tap delay channel, with the complex amplitude, $\beta_k$, depending on the target only {\cite{bao2019precoding,baquero2019full}},  and the delay  depending on the transmit and receive antenna indices, the target and the subcarrier frequency, i.e., the channel is  frequency selective. 

\vspace{-2mm} 
\subsection{Assumptions}
In this paper we make the following assumptions:
 
 \begin{enumerate}
 \item The length of the CP is larger than the maximum
roundtrip delay to the target, so that inter-symbol
interference  can be avoided during demodulation.
\item Secondary reflections from and to the target are attenuated and contribute noise only at the receiver.
\item \label{assump:coef}
{The coefficients $\beta_k$  are constant over $N_p$ OFDM symbols, and they do not need to be known.} Later we will discuss how we can estimate them during the sensing process. 
\item 
The noise terms $u_i(m,t)$ are independent identically distributed (i.i.d.)  white Gaussian noise processes with zero mean and variance $\sigma_r^2$ and do not depend on the frequency.
\item \label{assump:Doppler}
The OFDM signal bandwidth is much smaller than the carrier frequency, thus, within the same OFDM symbol, the phase shifts due to the Doppler effect are identical on all  subcarriers.
\item  \label{assump:range}
The duration of the OFDM symbol is  small enough to assume that the target range  is constant over $N_p$ OFDM symbols.
\item \label{assump:velo}
The Doppler frequency induced by the target is assumed to be constant over  $N_p$ OFDM symbols. 
 \end{enumerate}

\textcolor{black}{To see how realistic  assumptions \ref{assump:coef}, \ref{assump:range} and \ref{assump:velo} are, let us consider an OFDM system with subcarrier spacing of $90.909$kHz. The duration of each OFDM symbol is $12.375\mu s$ \cite{ofdm}. 
Thus, for $N_p = 256$ OFDM symbols, the duration is $3.168ms$.  With  high probability the targets will remain in the same angle, range and Doppler bin  during that time, which supports  assumptions \ref{assump:coef}, \ref{assump:range} and \ref{assump:velo}.}

%


\vspace{-2mm}
\subsection{Target Angle Estimation} \label{coarse}

Due to the existence of CP, the radar receive antennas can recover the precoded symbols via an $N_s$-point discrete Fourier transform (DFT). 
For simplicity, the rectangular pulse function and CP will be omitted in the following formulations.

Based on the above assumptions, the received signal of \eqref{receivedsig} can be written as
\begin{equation}
        y(m,i,t)  \approx \sum_{k=1}^{K}\sum_{n =1}^{N_t}  \beta_{k}  x(n,i,t-\tau^i_{kmn})e^{j2\pi f_{d_k}t}+ u_i(m,t)
        \label{receivedsig_new}
\end{equation}

Let us consider an observation interval equal to the duration of the  OFDM symbol duration.
Based on our assumptions,
the received signal of \eqref{receivedsig_new} \textcolor{black}{on the $i$-th subcarriers and in the $\mu$-th OFDM symbol equals
\begin{align}
        &y(m,i,t) = \sum_{k=1}^{K}  e^{-j2\pi mg_r\sin\theta_k\frac{f_c+i{\Delta f} }{c}}
      \nonumber \\
        &\times 
        \sum_{n=1}^{N_t} \beta_{k}d(n,i,\mu)
        e^{-j2\pi ng_t\sin\theta_k\frac{f_c+i{\Delta f} }{c}} \nonumber\\
        &\times
        e^{-j2\pi i{\Delta f} \frac{2R_k}{c}}
        e^{j2\pi \mu T_pf_{d_k}}e^{j2\pi i{\Delta f} t}+  u_i(m,t).
        \label{receivedsign_ext}
\end{align}
}
\textcolor{black}{Continuous time is typically expressed as  $t=\mu T_p+\tau$, where $\tau\in [0,T_p)$. When we sample in time, 
the sampling index corresponding to $\tau$  is usually referred to as \textit{fast time}, while the OFDM symbol index, $\mu$, is referred to as \textit{slow time}. 
Assumption (6)
allows us to separate range and Doppler effects. As  can be seen from Eq.\eqref{receivedsign_ext}, the phase shift due to range changes with respect to fast time, while the phase shift due to Doppler changes in slow time only.}
This makes it possible to measure range and Doppler independently.

\begin{figure}
     \centering
     \includegraphics[width = 7cm]{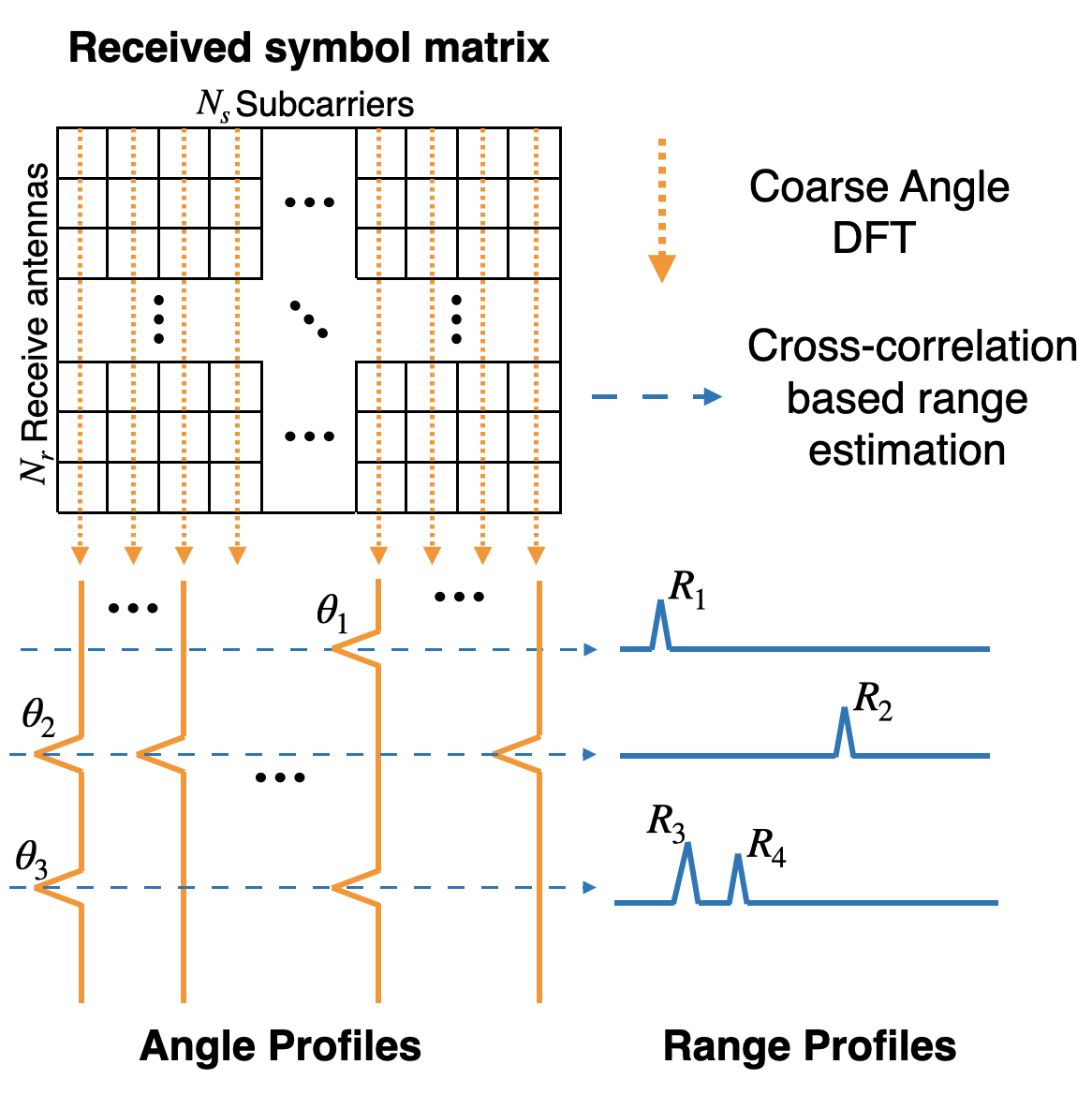}
     \caption{Estimation of angle and range in one OFDM symbol.}
     \label{fig:estimation}
     \vspace{-3mm}
\end{figure}

After sampling the received signal of duration $T_p$ and applying an $N_s$-point DFT on the received samples \cite{ofdm},
 \textcolor{black}{the fast time domain is transformed to the index of subcarriers and}
the symbol received by the $m$-the receive antennas equals 
\begin{align}
    d_{r}(m,i,\mu) =& \sum_{k=1}^{K}\sum_{n=1}^{N_t} \beta_{k} d(n,i,\mu)
        e^{-j2\pi(mg_r+ng_t)\sin\theta_k\frac{f_c+i{\Delta f} }{c}} \nonumber\\
        &\times
        e^{-j2\pi i{\Delta f} \frac{2R_k}{c}}e^{j2\pi \mu T_p f_{d_k}}
        + U(m,i,\mu).
        \label{prieq}
\end{align}
where $U(m,i,\mu)$ denotes the $N_s$-point DFT of the noise during the $\mu$-th OFDM symbol.
\textcolor{black}{
}
 Eq.\eqref{prieq} can be viewed  as
\begin{align}
    d_{r}(m,i,\mu) =\sum_{k=1}^{K} &A(k,i,\mu) e^{j \omega(k,i)m}+ U(m,i,\mu), 
    \label{fre_eq}
\end{align}
 for $m=0,...,N_r-1$, where
\begin{align}
    A(k,i,\mu)=& \sum_{n=1}^{N_t} \beta_{k} d(n,i,\mu)e^{-j2\pi ng_t\sin\theta_k\frac{f_c+i{\Delta f} }{c}}
    \nonumber\\&\times e^{-j2\pi i{\Delta f} \frac{2R_k}{c}}e^{j2\pi \mu T_p f_{d_k}}
    \label{ap1}
\end{align}
and
\begin{align}
    \omega(k,i) = -g_r\sin \theta_k \frac{f_c+i{\Delta f}}{c} 
    \label{omega}
\end{align}
\textcolor{black}{The symbols $d(n,i,\mu)$ are known to the DFRC receiver as this is a monostatic radar.}
\textcolor{black}{
One option to extract the target parameters is range-Doppler processing using DFT and cross-correlating the received signal at each range-Doppler bin with the transmit symbols, as performed in MIMO radar signal processing \cite{Tabrikian2008performance}. A computationally simpler, suboptimal approach is described next, and is also illustrated in 
Fig.~\ref{fig:estimation}.
} 
 
 Assuming that $N_r>K$ and for a fixed $i$ and $\mu$, the sequence $\{ d_{r}(m,i,\mu), m=0,...,N_r-1\}$ can be viewed as a sum of $K$ complex sinusoids with spatial frequencies $\omega(k,i)$ and complex amplitudes $A(k,i,\mu)$. 
Thus, upon  applying an $N_r$-point DFT on that sequence, we can get peaks at frequencies $\omega(k,i)$ (see Fig. \ref{fig:estimation}).
The resolution of the peaks will depend on the number of receive antennas, $N_r$.
Once the $\omega(k,i)$'s are estimated, the target angles can be computed as
\begin{align}
    \theta_k = \arcsin \left({- \frac{\omega(k,i)c}{g_r(f_c+i{\Delta f})}}\right)
    \label{angle}
\end{align}

\textcolor{black}{By repeating the angle DFT process on all subcarriers,  {thus exploiting frequency  diversity,} one can find all  occupied angle bins with high probability.} 


\textcolor{black}{We should note that in the above estimation, the complex amplitude $\beta_k$ does not need to be known, because the coarse angle estimation is along the receive array domain.}


\vspace{-2mm}
\subsection{Range and Doppler Estimation}\label{sec:Range&Doppler}

The amplitudes corresponding to  frequencies $\omega(k,i)$ contain known precoded symbols, estimated target angles, and unknown ranges and Doppler frequencies. 

Inside each angular bin, there may
be multiple targets. 
Suppose that there are $N_k$ targets corresponding to the $k$-th estimated direction $\theta_{k}$. Then, the corresponding amplitude can be expressed as
\begin{align}
    A(k,i,\mu) &= \sum_{n=1}^{N_t}    d(n,i,\mu)
        e^{-j2\pi ng_t\sin\theta_{k}\frac{f_c+i{\Delta f} }{c}}\nonumber\\
        &\quad\times\sum_{q=1}^{N_k} \beta_{kq} e^{-j2\pi i{\Delta f} \frac{2R_q}{c}} e^{j2\pi \mu T_p f_{d_q}}\nonumber\\
        &=   \sum_{q=1}^{N_k} \beta_{kq} A'(k,i,\mu) e^{-j2\pi  i{\Delta f} \frac{2R_q}{c}}e^{j2\pi \mu T_p f_{d_q}}
        \label{A}
\end{align}
where $\beta_{kq}$ is the coefficient of the $q$-th target at angle $\theta_k$ and 
\begin{equation}
    A'(k,i,\mu)=\sum_{n=1}^{N_t}  d(n,i,\mu)e^{-j2\pi ng_t\sin\theta_k\frac{f_c+i{\Delta f} }{c}} \label{Aprime}
\end{equation}


 Eq.  \eqref{A} can be written as
\begin{equation}
    A(k,i,\mu) = \sum_{q=1}^{N_k} \beta_{kq} e^{j2\pi \mu T_p f_{d_q}}A'(k,i,\mu)e^{j2\pi i\omega_r(q)}
    \label{eq: range_A}
\end{equation}
where
\begin{align}
    &\omega_r(q) = -{\Delta f} \frac{2R_q}{c}.
\end{align}
Let $\mathcal{A}(k,\ell,\mu)$ and $\mathcal{A}'(k,\ell,\mu)$ denote respectively the $N_s$-point DFT of $A(k,i,\mu)$ and $A'(k,i,\mu)$ {along  dimension $i$}.
One can see that $\mathcal{A}(k,\ell,\mu)$ is a weighted sum of shifted versions of $\mathcal{A}'(k,\ell,\mu)$, where  $\omega_r(q)$ are the shifts and
 $ \beta_q e^{j2\pi \mu T_p f_{d_q}}$ the weights. Since $A'(k,i,\mu)$ has already been estimated, the  shifts $\omega_r(q)$ can be measured based on the location of the peaks in the cross correlation of $\mathcal{A}'(k,\ell,\mu)$ and $\mathcal{A}(k,\ell,\mu)$. The peaks appear at indices
\begin{equation}
  l_q = \Big\lfloor\frac{2N_sR_q{\Delta f} }{c}\Big\rfloor, 
    \label{rangeind}
 \end{equation} \textcolor{black}{revealing the ranges of the targets that  fall in angle bin $\theta_k$, i.e., $R_q, i=1,...,N_k$. }

\textcolor{black}{
We should note that the obtained  range estimates use the full bandwidth, and thus have the maximum possible resolution. Errors in  angle estimation do not affect the range resolution, because
 angle estimation is conducted along the receive antenna index domain,  while  range estimation along the  subcarrier index domain. However, if the angle resolution is low and as a result fewer angles are estimated, some ranges may be missed. Later, in Section~\ref{sec:angle-range}, we will discuss how we can correct this potential problem by exploiting private subcarriers.}

For each estimated range, the corresponding Doppler frequency can be estimated by taking an $N_p$-point DFT  of the cross correlation peak values, i.e., $ \beta_q e^{j2\pi \mu T_p f_{d_q}}$, corresponding to  $N_p$ OFDM symbols, i.e., for $\mu=1,...,N_p$. The DFT  will contain peaks at indices
\begin{equation}
    p_q = \lfloor N_pT_pf_{d_q}\rfloor =\Big\lfloor\frac{2v_qf_cN_pT_p}{c}\Big\rfloor,  
    \label{veloind}
 \end{equation}
which provide the  targets Doppler and thus their velocities.
(see Fig. \ref{fig:estimation}).
By estimating the Doppler parameter  for each range peak, we can match the Doppler estimates with the  estimated ranges.
\textcolor{black}{Since the target ranges are estimated within certain angle bins, and the Doppler frequencies are estimated based on each range peak (see Fig.~\ref{fig:estimation}), the  angle-range-Doppler parameters of a given target are paired together.}
Again, the complex amplitudes $\beta_q$ do not need to be known, and they can actually be estimated based on the values of Doppler DFT peaks. 

\textcolor{black}{The proposed cross-correlation based method exploits the available angle information from the previous coarse angle estimation, thus avoiding searching over the entire parameter space.}
\textcolor{black}{We should note here that for the cross correlation peaks to be resolvable, the spectrum of $\mathcal{A}'(k,\ell,\mu)$ should be narrow around zero. Based on our experience with simulations this holds reasonably well, (see Section~\ref{simulation_section}.B)}.



For range estimation, the resolution is 
\begin{equation}
    R_{res} = \frac{c}{2N_s {\Delta f}} = \frac{c}{2B}
    \label{eq:range_res}
\end{equation}
where $B$ is the bandwidth of the OFDM waveforms. The maximum detectable range is determined by the spacing of subcarriers, i.e.,
\begin{equation}
    R_{max} = \frac{c}{2{\Delta f}}.
    \label{eq:range_max}
\end{equation}
Similarly, the resolution of the target velocity estimation and maximum detectable  velocity are
\begin{align}
    &v_{res} = \frac{c}{2 f_c N_p T_p}\\
    &v_{max} = \frac{c}{2 f_c T_p} 
    \label{eq:vel_max}
\end{align}
respectively. Note that the radial velocities could be both positive and negative, thus the maximum unambiguous velocity is half of the detectable velocity.

\textcolor{black}{
Based on \eqref{eq:range_max}, for a large unambiguous range  the subcarrier spacing should be as small as possible. On the other hand,  based on \eqref{eq:vel_max},  and since the OFDM symbol duration equals \textcolor{black}{$T_p =\frac{1}{\Delta f} + T_{cp}$, where $T_{cp}$ is the duration of the cyclic prefix,}
for higher detectable velocity
the subcarrier spacing should be  as large as possible. 
Thus in real world systems, the subcarrier spacing needs to be chosen carefully paying attention to the trade-off between the maximum unambiguous range and the maximum detectable velocity.
}

\section{Trading off communication rate for target estimation performance using private subcarriers}

\textcolor{black}{Angle estimation based on shared subcarriers relies on the aperture of physical array, i.e.,  $(N_r-1)g_r$. As a result, angle resolution is  limited by the size of the receive array. In the following, we will refer to those  angle estimates  as  \textit{coarse}.
\textcolor{black}{Low angle resolution may render some target angles unresolvable,  may result in erroneous angles.
Since the angle estimates are used to obtain target range and Doppler, the errors will propagate to those parameters as well.}} 

\textcolor{black}{
In this section we introduce the concept of private subcarriers that enables one to trade off  communication rate for better target estimation performance. A private subcarrier is exclusively assigned to  one transmit antenna, i.e., it  contains symbols from that  antenna only. Thus,  the use of private subcarriers decreases the overall communication rate. However, as it will be discussed next, the private subcarriers can be leveraged to improve sensing performance.}

Let us assume that we allocate   $M$ subcarriers as private, where $1\le M\le N_t$
and let  $\mathcal{M}$ denote the set of private subcarriers.
Each private subcarrier $i$ is uniquely assigned to transmit antenna $n_i$.
The remaining  $N_s-M$ subcarriers are assigned to transmit antennas   in a shared fashion.

Suppose that we obtain target estimates based on all (shared and private) subcarriers.
\textcolor{black}{The methodology for angle and range estimation presented in Section II still applies to the   system that uses both private and  shared subcarriers. This is because  \eqref{eq: range_A}
still holds on private subcarriers.} As we will show next, the VA will allow us to fine-tune those estimates.

Let there be no precoding on the private subcarriers,
and the sole data symbol on  private subcarrier $i$ be denoted by $Q(n_i,i)$.
In order to keep the same power level on private subcarriers as on the shared ones, $Q(n_i,i)$ is appropriately scaled.
The transmitted symbol vector on the $i$-th subcarrier is
\begin{equation}
    \mathbf{d}_i =  Q(n_i,i).
    \label{private_trans_sym}
\end{equation}

From \eqref{prieq}, {omitting the noise term} and  performing  element-wise division by the known data symbol that was transmitted on private subcarrier $i$, the received symbol on the same subcarrier becomes 
\begin{align}
    d'_r(m,i,\mu)= &\sum_{k=1}^{K} \beta_{k} e^{-j2\pi (mg_r + n_i g_t) \frac{\sin{\theta_k}}{\lambda_i}} \nonumber\\
    \times &e^{-j2\pi i{\Delta f} \frac{2R_k}{c}} e^{j2\pi \mu T_p f_{d_k}}
    , \quad i\in\mathcal{M}. 
    \label{eq:pri_symbols}
\end{align}

\textcolor{black}{
For simplicity, here we let $\mathcal{M}=\{0,1,\dots,M-1\}$ and set $n_i = i$.
}


\textcolor{black}{Let us define the transmit and receive steering vectors corresponding to the $i$-th subcarrier as 
\begin{eqnarray}
    \mathbf{a}_r(\theta,i) = [1, e^{-j2\pi g_r  \frac{\sin{\theta} }{\lambda_i}}, ..., e^{-j2\pi(N_r-1)g_r  \frac{\sin{\theta} }{\lambda_i}}]^T \label{eq:receive_st}\\
    \mathbf{a}_t(\theta,i) = [1, e^{-j2\pi g_t  \frac{\sin{\theta} }{\lambda_i}}, ..., e^{-j2\pi(N_t-1)g_t  \frac{\sin{\theta} }{\lambda_i}}]^T
\end{eqnarray}
respectively,  where $\lambda_i = \frac{f_c+i{\Delta f}}{c}$.
Then we define a matrix containing all the transmit steering vectors on the private subcarrier as
\begin{equation}
    \mathbf{A}^p_t(\theta) = [\mathbf{a}_t(\theta,0),\dots, \mathbf{a}_t(\theta,M-1)]
\end{equation}
Since only one antenna transmits on each private subcarrier, the transmit steering vector corresponding to the private subcarriers can be written as
 \begin{equation}
      \mathbf{a}^p_t(\theta) = [1, e^{-j2\pi g_t  \frac{\sin{\theta} }{\lambda_1}}, ..., e^{-j2\pi(M-1)g_t  \frac{\sin{\theta} }{\lambda_M}}]^T
 \end{equation}}

\textcolor{black}{In the following, in order to make the presentation of the idea easier, and the connection a virtual array more obvious, we will assume that the private subcarriers are closely spaced. In that case, we can
 approximate $\frac{f_c+i{\Delta f}}{c} \approx \frac{f_c}{c} = \lambda_0, i\in\mathcal{M}$,
and will omit the subcarrier index in the receive steering vector notation.
Later in this section (see Remark 3) we  will explain how  the same conclusions can be drawn without making the aforementioned approximation, i.e., taking the private subcarriers in any location. }

\subsection{Refining the  angle estimates}\label{refine}

Let $i\in \cal M$, and 
define  $\Tilde{\mathbf{a}}_t(\theta_k,R_k)\in \mathbb{C}^{M\times 1} $ as
\begin{equation}
    \Tilde{\mathbf{a}}_t(\theta_k,R_k) \buildrel \triangle \over = \mathbf{a}^p_t(\theta_k) \odot \mathbf{b}(R_k)
\end{equation}
where $\odot$ denotes Hadamard product and
\begin{align}
    \mathbf{b}(R_k) = [1,e^{-j2\pi {\Delta f} \frac{2R_k}{c}},\cdots, e^{-j2\pi (M-1){\Delta f} \frac{2R_k}{c}}]^T
\end{align}

By stacking the  symbols of   \eqref{eq:pri_symbols} for all $i\in\mathcal{M}$ in  a column vector, i.e., $\mathbf{z}_m \in \mathbb{C}^{M\times 1}$, we have that
\begin{equation}
    \mathbf{z}_m = \sum_{k=1}^{N_k}\beta_k e^{j2\pi \mu T_p f_{d_k}}e^{-j2\pi mg_r\frac{\sin{\theta_k}}{\lambda_0}} {\Tilde{\mathbf{a}}}_t(\theta_k,R_k) .
\end{equation}
By stacking the resulting vectors from all  receive antennas into a long vector, i.e., $\mathbf{z}$, we get
\begin{align}
    \mathbf{z} = \sum_{k=1}^{N_k} \beta_k e^{j2\pi \mu T_p f_{d_k}} \mathbf{a}_r(\theta_k) \otimes {\Tilde{\mathbf a}}_t(\theta_k,R_k),
    \label{virarr}
\end{align}
{where  $\mathbf{a}_r(\theta_k)$} is the receive steering vector of \textcolor{black}{\eqref{eq:receive_st}}  but here we omit the index $i$, and 
where $\otimes$ is the Kronecker product.

\textcolor{black}{
In the above expression, the   signal received on all private subcarriers of all receive antennas, i.e., ${\bf z}$,
can be effectively viewed as the response of a  VA
\cite{Li2007MIMO}. The VA has an aperture $M$ times the aperture of the physical receive array, and thus can enable better angle resolution. Unlike typical VAs, however, its steering vector, i.e., $\mathbf{a}_r(\theta_k) \otimes {\Tilde{\mathbf a}}_t(\theta_k,R_k)$, depends on target range as well as angle.
}

\begin{figure}
    \centering
    \includegraphics[width = 7.5cm]{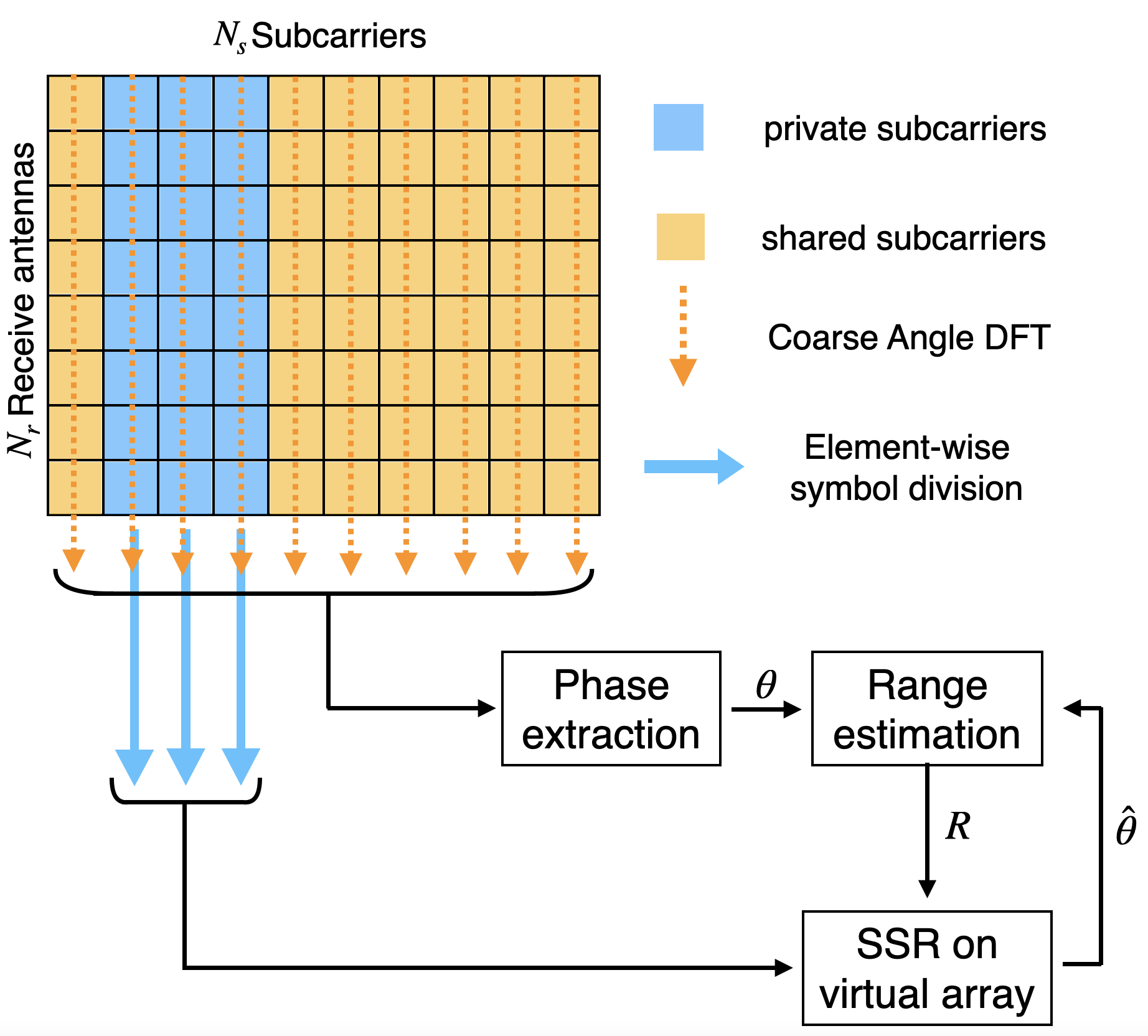}
    \caption{Target estimation with private subcarriers in an SS-OFDM DFRC system.}
    \label{fig:virtual_array}
    \vspace{-3mm}
\end{figure}

Angle estimation proceeds as follows.
 Let $\{R_{1},R_{2},...,R_N\}$ be the $N$ previously obtained range estimates \textcolor{black}{(see section~\ref{sec:Range&Doppler})}.
By discretizing the angle space on a grid of size $N_a$, i.e., $\{\tilde \theta_1, ..., \tilde \theta_{N_a}\}$,   \eqref{virarr} can be expressed as
\begin{eqnarray}
    \mathbf{z} &=& [\mathbf{z}_{11},\mathbf{z}_{12},...,\mathbf{z}_{N_aN}]
    \begin{bmatrix}
    \Tilde{\beta}_{11},\\...,\\\Tilde{\beta}_{N_aN}
    \end{bmatrix}\nonumber\\
    &=& [\mathbf{z}_{11},\mathbf{z}_{12},...,\mathbf{z}_{N_aN}]{\tilde{\beta}}
    \label{base_matrix}
\end{eqnarray}
where $\tilde \beta_{ij}$ is non zero if there is a target at range $R_{j}$ and angle $\tilde \theta_i$, and
\begin{equation}
    \mathbf{z}_{ij} = \mathbf{a}_r(\tilde \theta_i)\otimes({\Tilde{\mathbf a}}_t(\tilde \theta_i) \odot \mathbf{b}(R_{k_j}))
    \label{base_vector}
\end{equation}
is the dictionary element for $i = 1,2,...,N_a$ and $j = 1,2,...,N$. \textcolor{black}{When the target angle space \textcolor{black}{within the considered range bins is sparse,} vector ${\tilde \beta}$ is sparse and under certain conditions can be estimated  \cite{Kalogerias2013Matrix}
%
%
via $\ell_1$-norm minimization \cite{baraniuk2007compressive}. Its support  provides the angle space grid points that are the closest to the target angles, yielding
refined target angle estimates.} 
\textcolor{black}{We should note that, upon using the previously estimated ranges to construct the base matrix, the refined angle estimates are naturally paired with the range estimates.}
\textcolor{black}{The entire estimation process is illustrated in Fig.~\ref{fig:virtual_array}. }


    \medskip
    
\noindent\textbf{Remark 1:} {The aperture of the virtual array is determined by the number of private subcarriers, $M$,  and is maximum when $M=N_t$. Given that the use of private subcarriers comes at a cost of communication rate, there is no reason for $M$ to be  larger than $N_t$.}

\smallskip
 \noindent\textbf{Remark 2:} One may argue that the SSR problem could be constructed without using the available range estimates, by discretizing the entire range space as well as the angle space and forming an overcomplete problem  for both range and angle estimation.
However, such approach would have the following shortcomings. The  entire range and angle space may not be sparse enough for SSR to work well; it is only  within specific range bins that the target space can be assumed to be sparse.
Also, such approach would not exploit available information. {With a large number of subcarriers in the OFDM system, the range estimates, obtained {based on the coarse angle estimates}  are of high resolution as they use all subcarriers (although some ranges may have been missed due to the low resolution of the phase estimates).  }
 Another reason against such approach would be  the high dimension of  $\tilde \beta$, which would increase the SSR complexity.
 Therefore, the proposed iterative approach that uses a combination of private and shared subcarriers achieves not only high sensing performance but also avoids unrealistic assumptions and reduces estimation complexity.
 
 \smallskip
 \noindent\textbf{Remark 3:} 
\textcolor{black}{Let us avoid assumption  $\lambda_i\approx \lambda_0$ for $i \in \cal M$. {In that case, the receive steering vector   depends on  subcarrier index.} Let us stack the  symbols received on   all  receive antennas  (see \eqref{eq:pri_symbols}) in vector $\mathbf{z}_i \in \mathbb{C}^{N_r\times 1}$, i.e., 
\begin{align}
    \mathbf{z}_i &= \sum_{k=1}^{K}\beta_{k} e^{j2\pi \mu T_p f_{d_k}}
    e^{-j2\pi n_ig_t\frac{\sin{\theta_k}}{\lambda_i}} e^{-j2\pi i{\Delta f} \frac{2R_k}{c}} {\mathbf{a}}_r(\theta_k,i) .
    \label{eq: z_new}
\end{align}
By stacking the resulting vectors in \eqref{eq: z_new} from all  private subcarriers into a long vector, i.e.,
    $\tilde{\mathbf{z}} =[ \mathbf{z}_0, ..., \mathbf{z}_{M-1}]^T,
    \label{virarr_new}$
we get a $MN_r \times 1$  long vector which is similar to \eqref{virarr}. Then by formulating a base matrix with dictionary element sharing a same structure of \eqref{eq: z_new}, one can apply the same SSR method to obtain refined target angle estimates.}

\subsection{Angle and range iterative estimation}\label{sec:angle-range}

\textcolor{black}{After obtaining the  high resolution angle estimates by solving the SSR problem of \eqref{base_matrix},  we can revisit   \eqref{A}-\eqref{veloind} to re-evaluate the ranges. This process may reveal ranges that were previously missed.}


When using all (private and shared) subcarriers, the obtained  range estimates use full bandwidth, and thus have the maximum possible resolution. Errors in  angle estimation do not affect the range resolution, because
 angle estimation is conducted along the receive antenna index domain,  while  range estimation along the  subcarrier index domain.
{However, since we first separate targets into different angle bins, if due to low angle resolution some angles are not resolvable and  appear in the same angle bin,  the ranges of the targets corresponding to that angle bin will be correct but cannot be matched  with the exact angles. Also,  if in \eqref{prieq}, for some $k$,  it holds that $A(k,i,\mu)\approx 0$  for all subcarriers $i$, (this would happen with small probability), then the corresponding frequency will be missed, and as a result, the corresponding ranges will be missed.} 
Still, using those range estimates to construct the basis matrix of the SSR  problem we can high resolution angle estimates.

Based on the above, we propose an iterative angle-range estimation algorithm as summarized in Algorithm~\ref{alg:angle-range} (see also Fig.~\ref{fig:virtual_array}).
\textcolor{black}{
The iteration is initialized with the  angle estimates obtained on all subcarriers. Subsequently, in Step 2, the ranges are computed via    \eqref{rangeind} \textcolor{black}{based on all subcarriers} and using the  angles estimates  from the previous step. In Step 3, using the private subcarriers and the previously obtained range estimates, an SSR problem is formulated based on   \eqref{base_matrix}.  The solution of the SSR problem provides finer resolution angle estimates.  \textcolor{black}{Steps 2 and 3 are executed several times (typically 2-3) until no new targets are found.}}


We should note that the angle-range iterative estimation can be done using  one OFDM symbol only, while Doppler estimation requires $N_p$ OFDM symbols and is carried out in the slow time domain. 
By the time the echoes of $N_p$ OFDM symbols have been received,  the occupied angle and range bins would have already been estimated. Then, 
the Doppler frequencies can be estimated within the occupied angle bins based on the coefficients of the ranges peaks (see \eqref{veloind}).

%


\textcolor{black}{The proposed iterative estimation approach bears some similarity to the  method of \cite{Xu2015Joint} proposed for the problem of angle-range dependent virtual array,  arising when different antennas transmit on different subcarrier frequencies, although   our scenario faces more complicated coupling. 
Both methods first coarsely estimate the target parameters then use the results to refine estimation. 
The difference is that here we use the entire virtual array to estimate angle, thus enjoying  higher resolution, while \cite{Xu2015Joint}  uses the receive array only. Also, our method can recover targets unresolved in the coarse estimation, and the range estimates that we use for angle estimation are always at maximum resolution. Thus the results of our method are refined angle and range estimates, while the result in \cite{Xu2015Joint} are only refined range estimates.}

\begin{algorithm}[]
\SetAlgoVlined
\DontPrintSemicolon
Step 1: Obtain coarse angle estimates via \eqref{angle} on all subcarriers\;
\Repeat{
no changes are made in SSR results
}
{Step 2:  Obtain range estimates based on the obtained angle estimates, via \eqref{rangeind} \;
Step 3: Formulate and solve the SSR problem of \eqref{base_matrix} based on  the obtained range estimates, to obtain refined angle estimates\;}
\caption{Angle-Range iterative estimation}
\label{alg:angle-range}

\end{algorithm}

 \section{Communication System Model}
\textcolor{black}{Let us consider a communication receiver with $N_c$ antennas, spaced apart by ${d}_r$. 
The $i$-th subcarrier, between the $n$-th transmit antennas and the $\ell$-th receive antenna, undergoes an effect that can be modeled as propagation through a multiple tap delay channel. The first 
 delay $\tau^i_{n \ell}$, is due to the direct path from the radar transmitters, i.e., 
\begin{equation}
    \tau^i_{n \ell}=R_c/c + \ell d_r sin\theta/{\lambda_i}+ n g_t sin\phi/{\lambda_i},
\end{equation}
where 
$\theta$ is the angle of the communication system from the point of view of the radar,  $\phi$ is the angle of the radar from the point of view of the communication system,  and  $R_c$ is the distance between the two systems. The corresponding coefficient is denoted as $ \beta$.
The remaining tap delays are due to reflections of the radar transmissions and their contribution to the channel can be expressed as an additive term.
The channel frequency response corresponding to the $i$-th subcarrier can be expressed as 
\begin{eqnarray}
    {\bf H}_i=&\beta e^{-j2\pi i{\Delta f} R_c/c} {\bf a}_t(\theta,i)  {\bf a}^T_r(\phi,i) +\nonumber \\
    &\sum_k c_k {\bf a}_t(\theta_k,i)  {\bf a}^T_r(\phi_k,i) 
    \label{eq:channel}
\end{eqnarray}
where {$\mathbf{a}_t(\theta,i) = [1, e^{-j2\pi g_t  \frac{\sin{\theta} }{\lambda_i}}, ..., e^{-j2\pi(N_t-1)g_t  \frac{\sin{\theta} }{\lambda_i}}]^T$ is the transmit steering vector for angle $\theta$ on the $i$-th subcarrier and,}
$\theta_k,\phi_k$ are angles of departure and incidence related to the various scatterers, and $c_k$ the corresponding coefficients.
}

%

Assuming that the channel spread is smaller than the CP length, 
and due to the narrow bandwidth of each subcarrier in the OFDM system, the effect from a frequency selective fading channel between transmit antenna and receive antenna can be mitigated.



{On the shared subcarriers,}
 the symbols   across all receive antennas can be expressed as
 \begin{equation}
     \mathbf{r}_i = {\mathbf{H}}_i{\mathbf{d}}_i + \mathbf{u}_i, \ i = 0,...,N_s-1,
     \label{comm_sym_r}
 \end{equation}
 where $i$ is the subcarrier index,   $\mathbf{d}_i$ is the $i$-th column of $\d$ 
and 
$\mathbf{u}_i \in\mathbb{C}^{N_c\times 1}$ for $i = 0,...,N_s-1$ represents the measurement noise on the $i$-th carrier, which is assumed to be  white, Gaussian with zero mean and covariance $\sigma_c^2 {\bf I}$.
\textcolor{black}{Assuming that the communication receive array has more antennas than the radar transmit array, i.e., $N_c > N_t$, and  knows the precoding matrix and the constellation diagram used at the transmitter,}  {the data symbol vector can be estimated via  least-squares, i.e.,
\begin{equation}
    \begin{split}
        &\arg \min_{\mathbf{q}} ||\mathbf{r}_i - \mathbf{H}_i\mathbf{P}\mathbf{q}||_2^2\\
    \end{split}
    \label{MLestimate}
\end{equation}
}
The above assumes that the receiver has knowledge of the communication channel, which can be obtained via the transmission of pilots. 

\textcolor{black}{On the private subcarriers, the symbol recovery problem is slightly different, as there is no precoding matrix.
If the communication receiver knows the  location of the private subcarriers and the corresponding antennas, then, based on the private  subcarriers, the symbols can be decoded by finding the element in the constellation diagram, which after multiplication by the corresponding channel vector yields the smallest distance to the received symbol vector. } 
\textcolor{black}{The receiver can determine who are the  private subcarriers by  leveraging the fact that  the transmitted data symbol vector on private subcarriers is naturally $1$-sparse. The private subcarriers can be identified by solving a  least-squares problem without precoding on each subcarrier, and checking for   $1$-sparseness in the result.}

After the decoding process, the original information can be extracted from the recovered data symbols. By applying the same process to every subcarrier and every OFDM symbol,   all transmitted symbols can be recovered. Compared with an OFDM communication system with the same modulation scheme but without subcarrier sharing, the proposed  system  increases the number of information bits transmitted in  one period by a factor of up to $N_t$.

\textcolor{black}{
While the use of private subcarriers enables  higher angular resolution for sensing,  it comes  at a cost of reduced number of shared subcarriers. The corresponding  loss in the communication bit rate is $\frac{M-1}{N_s} \%$.
If $N_s >> M$, the loss from private subcarriers is small as compared to the aforementioned advantages.
}


\section{Precoder Design}\label{sec:design}

The precoder is obtained by optimizing a function of the beampattern error and the SNR at the communication receiver.

Let $\mathbf{F}\in \mathbb{C}^{N_s \times N_s}$ denote the inverse Fourier transform matrix.
The baseband transmitted signal corresponding to the $\mu$-th OFDM symbol can be expressed in matrix form as
\begin{equation}
    {\mathbf X} = \mathbf D{\mathbf F} = \p\d{\mathbf F}.
    \label{matform}
\end{equation}
The  $j$-th column of the baseband signal matrix, or otherwise, the $j$-th snapshot, equals
\begin{equation}
    \x_j = \sum_{i=0}^{N_s-1} \p\mathbf{d}_i F(i,j).
\end{equation}
where $F(i,j)$
is the element on the $i$-th row and $j$-th column of $\f$. 
The  array output towards direction $\theta$ at the $j$-th snapshot is
\begin{equation}
    y_j(\theta) = \sum_{i=0}^{N_s-1}{\mathbf a}_t^T(\theta,i)\p\mathbf{d}_i F(i,j),
\end{equation}
and the  transmitted power  equals
\begin{eqnarray}
    \hat{p}(\theta) &=& \mathbb{E}\{y_j(\theta)y_j(\theta)^*\} = \frac{1}{N_s}\sum_{j=0}^{N_s-1}y_j(\theta)y_j(\theta)^*\nonumber\\
    &=&  \frac{1}{N_s}\sum_{i=0}^{N_s-1}\sum_{j=0}^{N_s-1}{\mathbf a}_t^T(\theta,i)\p\mathbf{d}_i F(i,j)F^H(i,j)\nonumber\\
    && \qquad\qquad\qquad\qquad \times  \mathbf{d}^H_i\p^H{\mathbf a}_t^*(\theta,i)\nonumber\\
    &=& \frac{1}{N_s}\sum_{i=0}^{N_s-1}{\mathbf a}_t^T(\theta,i)\p\mathbf{R}_i\p^H{\mathbf a}_t^*(\theta,i)
\end{eqnarray}
where 
${\mathbf a}_t^T(\theta,i)$ is the transmit steering vector corresponding to the $i$-th subcarrier, 
$\sum_{j=0}^{N_s-1}F(i,j)F^H(i,j) = 1 \ \text{for}\ i = 0,1,\dots,N_s-1$ and
$\mathbf{R}_i=\mathbf{q}_i\mathbf{q}_i^H$ can be viewed as the covariance matrix of the $i$-th original symbol vector $\mathbf{q}_i$ (the $i$-th column of matrix ${\bf Q}$), which is white. 

Let $p(\theta)$ be 
the desired power of the transmitted signal towards direction $\theta$. 
The  beampattern error with respect to the desired beampattern $p(\theta)$, equals

 \begin{align}
    \sum_{g=1}^{G}[p(\theta_g)-&\hat{p}(\theta_g)]^2 = \\\nonumber
    &\|\mathbf{p}- \sum_{i=0}^{N_s-1}\text{diag}\{ {\mathbf A}_i^T\p\mathbf{R}_i\p^H\mathbf{A}_i^*\}\|_2^2
\end{align}
where $\theta_g$ are discretized angles in $[-\pi/2,\pi/2]$ on a grid of $G$ points.
$\mathbf{A}_i = [\mathbf{a}_t(\theta_1,i),...,\mathbf{a}_t(\theta_G,i)]\in\mathbb{C}^{N_t\times G}$ is the transmit steering matrix on the $i$-th subcarrier.

\subsection{SNR at the communication receiver}
 
From \eqref{comm_sym_r}, the power of received signal from all $N_s$ subcarriers in the $\mu$-th OFDM symbol can be expressed as
\begin{align}
    \text{P}_r &= \mathbb{E}\{ tr[{\bf r}_i {\bf r}^H_i  ]\}=
    \frac{1}{N_s}\sum_{i=0}^{N_s-1} tr[ \h_i \mathbf{d}_i \mathbf{d}^H_i \h^H_i],
    \label{power1}\\
&= \frac{1}{N_s}\sum_{i=0}^{N_s-1}  tr[\h_i  \p \mathbf{R}_i \p^H  \h^H_i].
    \label{power2}
\end{align}
where $tr[\cdot]$ refers to the trace of a matrix.
Recall that the power of communication noise is $\sigma_c^2$, thus the SNR at the communication receiver equals
\begin{align}
     \text{SNR} =&\frac{\sum_{i=0}^{N_s-1} tr[\h_i  \p \mathbf{R}_i \p^H  \h^H_i]}{N_s\sigma_c^2}.
     \label{eq:snr}
\end{align}

\subsection{The sensing-communications co-design problem}

Let us consider the loss function
\begin{align}
    \mathcal{L}(\p) &\triangleq 
   \alpha_b\sum_{g=1}^G \gamma_i [p(\theta_g) - \hat{p}(\theta_g)]^2  \nonumber\\
    &+ \alpha_{snr}10\text{log}_{10}(\frac{N_s\sigma_c^2}{\sum_{i=0}^{N_s-1} tr[\h_i  \p \mathbf{R}_i \p^H  \h^H_i]}).
    \label{eq:learning_loss}    
\end{align}
where 
$\alpha_b$ and $\alpha_{snr}$ are cost parameters,  respectively reflecting the relative importance of the beampattern error (first term)  and the inverse SNR (second term); 
$\gamma_i$ are weights that control the importance of beampattern error 
from the $i$-th angle.

\textcolor{black}{By adjusting the $\alpha_b$ and $\alpha_{snr}$, one can put more emphasize on approximating a desired beampattern for radar purposes, or maximizing the SNR for communication purposes. Since the beampattern error is accumulated over all the angle grids, $\alpha_b$ should be much smaller than $\alpha_{snr}$. 
%
}
%
%







\section{Numerical Results}\label{simulation_section}

 
In this section, we demonstrate the sensing and communication performance of the proposed \textcolor{black}{SS-DFRC} system  via simulations. 
The data symbols were modulated by quadrature phase shift keying (QPSK). {The average power of the symbols on the shared subcarriers is set to  ensure the same power level on all subcarriers.} 
The system parameters were taken to be consistent with those in other studies \cite{ofdm,Liu2017OFDMDFRC,Dokhanchi2018OFDMDFRC}, and are shown in Table~\ref{table:system_parameters}.
\textcolor{black}{
The channels were simulated based on \eqref{eq:channel}, where the coefficients, \textcolor{black}{$\beta_{k}$ were generated as  complex random variables with mean \textcolor{black}{$0.1$} and variance  $0.01$. }}

\begin{table}[!h]
\caption{System Parameters}
\label{table:system_parameters}
\centering
\resizebox{75mm}{18mm}{
\begin{tabular}{ |c||c|c|  }
 \hline
 Parameter & Symbol & Value\\
 \hline
 Center frequency   & $f_c$    &24GHz\\
 Subcarrier spacing &   ${\Delta f} $  & 0.25MHz\\
 Duration of  OFDM symbol & $T_p$ & 5$\mu$s\\
 Number of subcarriers & $N_s$ & 512\\
 Number of OFDM symbols & $N_p$ & 256\\
 Number of radar receive antennas & $N_r$ & 32\\
 Number of communication receive antennas & $N_c$ & 64\\
 Receive antenna spacing distance & $g_r$ & 0.5$\lambda$\\
 Transmit antenna spacing distance & $g_t$ & 0.5$\lambda$ \\
 \hline
\end{tabular}}
\label{table:parameters}
\end{table}

\subsection{Beampattern design} 

In this section we  obtain the  precoder that minimizes the loss function of \eqref{eq:learning_loss}, and  show the corresponding SNR gain.


The desirable  beam power profile was set to be $0$ everywhere except over the angle range $[-52,-37]$ degrees, corresponding to the region of interest for the radar, and  over the angle range $[29,31]$ degrees, corresponding to the region where the communication receiver is located; over those angles ranges  the beam profile was set to $1$.
The weights in the total loss function  of \eqref{eq:learning_loss} were take as $\alpha_b = 1\times 10^{-4}$, and $\alpha_{snr}=0.8$.

To solve \eqref{eq:learning_loss}
we used the Adam stochastic optimizer \cite{Kingma2014adam}  with different learning rates, starting at  $0.02$, and $N_{step}=150$ steps of iterations. {Note that the learning rates here only relate to the step length. Different learn rates, i.e., step lengths help us avoid local optima. A large number of executed steps ensures  stable convergence. 
}


\begin{figure}
{\small
    \centering
    \includegraphics[width = 8cm]{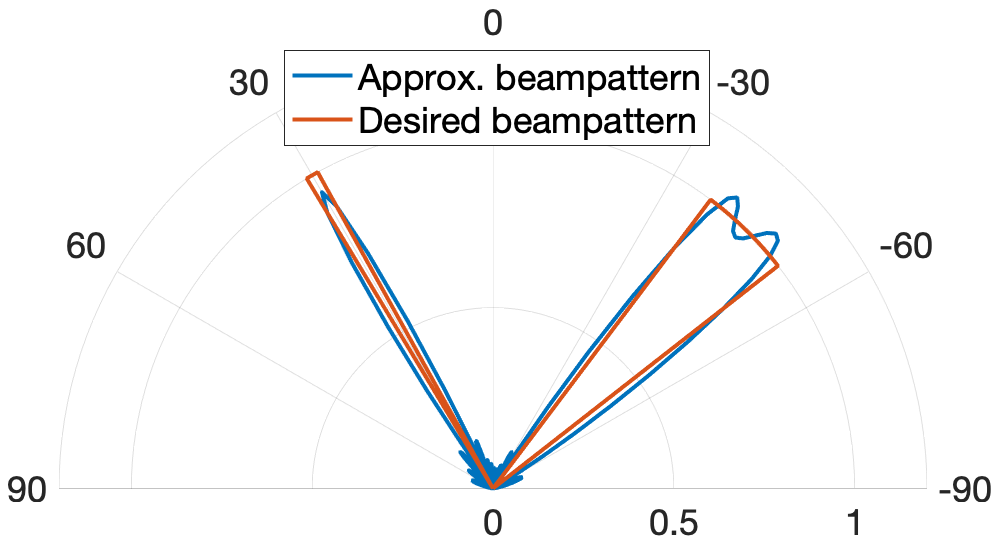}
    \caption{Beampattern based on $N_{t}=16$ antennas.}
    \label{fig:beampattern}
    }
    \vspace{-3mm}
\end{figure}


\begin{figure}
{\small
    \centering
    \includegraphics[width = 6.5cm]{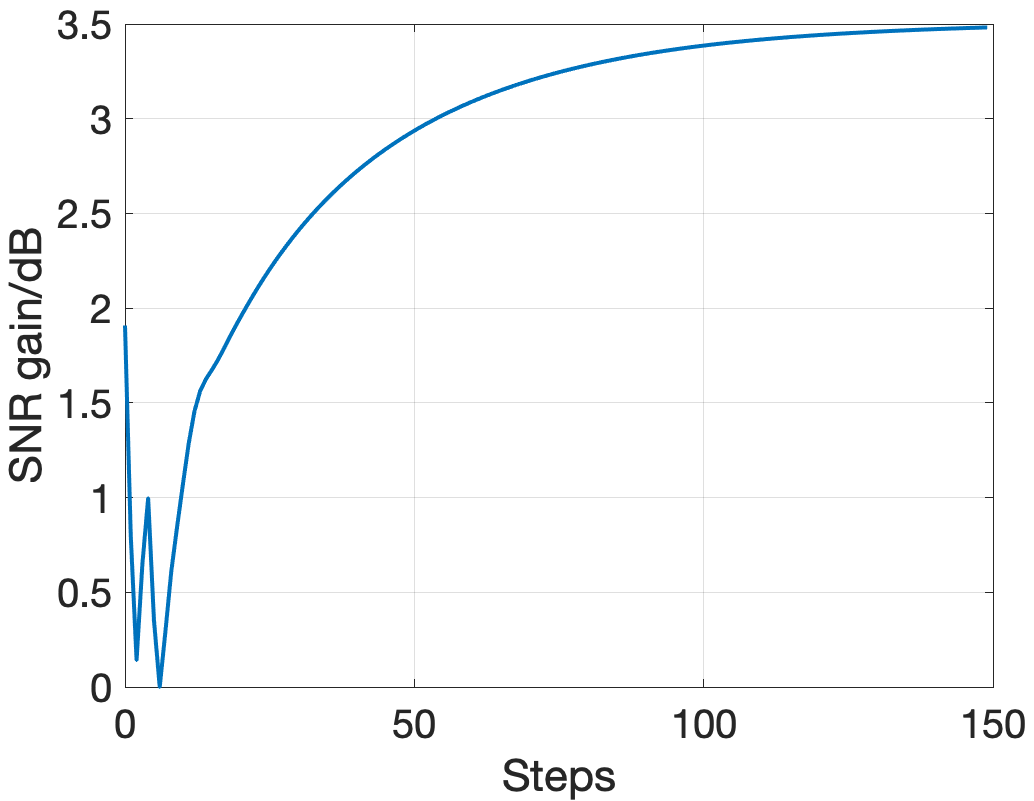}
    \caption{SNR gain with $16$ antennas.}
    \label{fig:snr}
    }
    \vspace{-3mm}
\end{figure}

\begin{figure}
{\small
    \centering
    \includegraphics[width = 6.5cm]{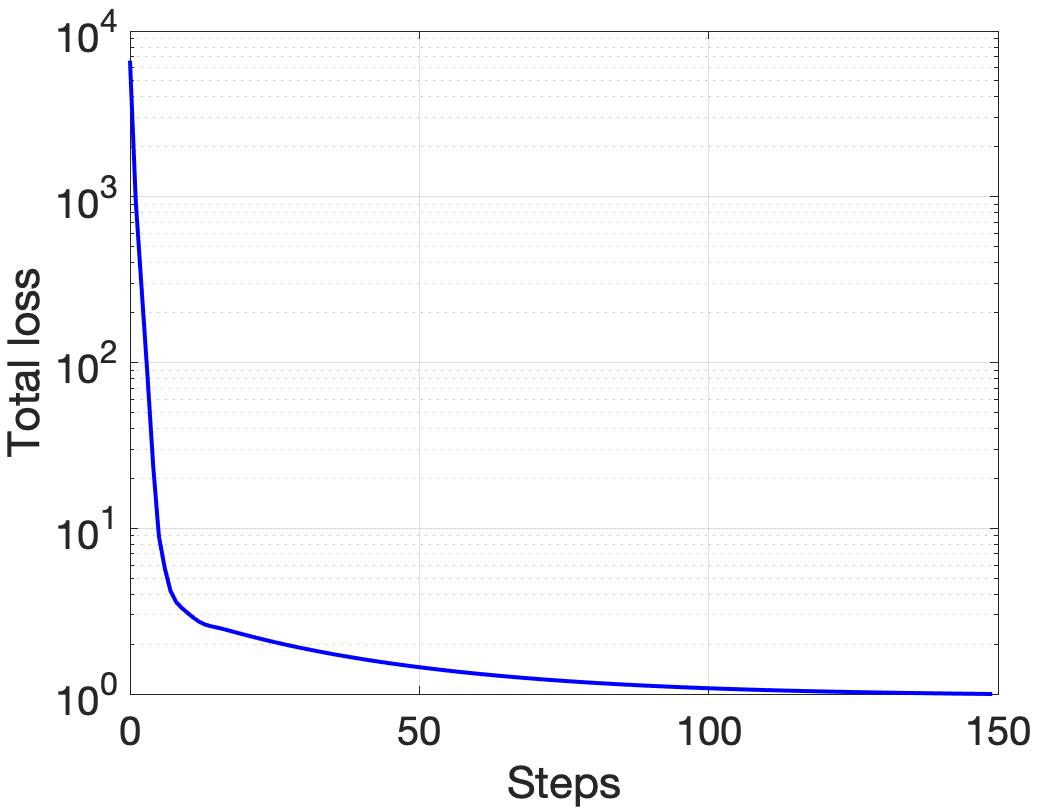}
    \caption{The loss curve during optimization.}
    \label{fig:convergence}
    }
\end{figure}

The optimum precoding matrix simultaneously approximates the beampattern to a desired one and  maximizes the SNR at the communication receiver.
Fig.~\ref{fig:beampattern}, shows the designed beampattern when having $N_t = 16$ antennas.
\textcolor{black}{
Fig.~\ref{fig:snr} shows the change of SNR during the iteration with respect to the initial value.
One can see that the SNR gain first grows and drops rapidly in the first $10$ epochs, and then increases slowly and reaches convergence.
The rapid increase in SNR gain is due to the large weight, $\alpha_{snr}$, as compared with $\alpha_{b}$. The subsequent drop in SNR  is due to the optimization with respect to  the precoding matrix, which aim to reduce the beampattern error.}
In general, when the beampattern error dominates  the total loss, the method will choose to optimize with respect to  $\p$ to improve the beampattern error. Once the  beampattern error becomes smaller than the loss due to SNR, the model chooses to optimize $\p$  to improve the communication performance at the cost of increasing the beampattern error. 
The corresponding learning curve is shown in Fig.~\ref{fig:convergence}.
We should note  that the loss function \eqref{eq:learning_loss} could have negative values. Thus for the sake of plotting,  an offset s added to the loss function so that the minimum is $1$.

\subsection{Sensing performance}

We consider $4$ point targets in the far field of the radar array, at angles, ranges, and velocities  as shown in Table~\ref{radar}. 
{We should note that based on the aperture of the receive array of $32$ antennas, used in this experiment, 
these targets fall in $2$ different angles bins;  each bin has $2$ targets at different ranges. }

The SNR  is set to be $15$dB, and the channel coefficients,  $\beta_{k}$, in \eqref{prieq} are assumed to be known.
In this experiment, the transmitter has $8$ antennas and the corresponding precoding matrix $\p$ is found by optimizing the criterion of \eqref{eq:learning_loss}.
Different number of  private subcarriers, $M$, is considered, starting  with $M=8$.

\begin{table}[]
\caption{Targets Parameters}
\centering
{
\begin{tabular}{ |c|c|c| }
 \hline
Angles & Ranges & Velocities \\
\hline
$-43\degree$ & $50m$ & $13m$/s\\ 
$-43\degree$ & $80m$ & $20m$/s\\ 
$-46\degree$ & $45m$ & $-10m$/s\\ 
$-48\degree$ & $100m$ & $10m$/s\\ 
\hline
\end{tabular}}
\label{radar}
\end{table}


\begin{figure}
{\small
    \centering
    \includegraphics[width = 6.5cm]{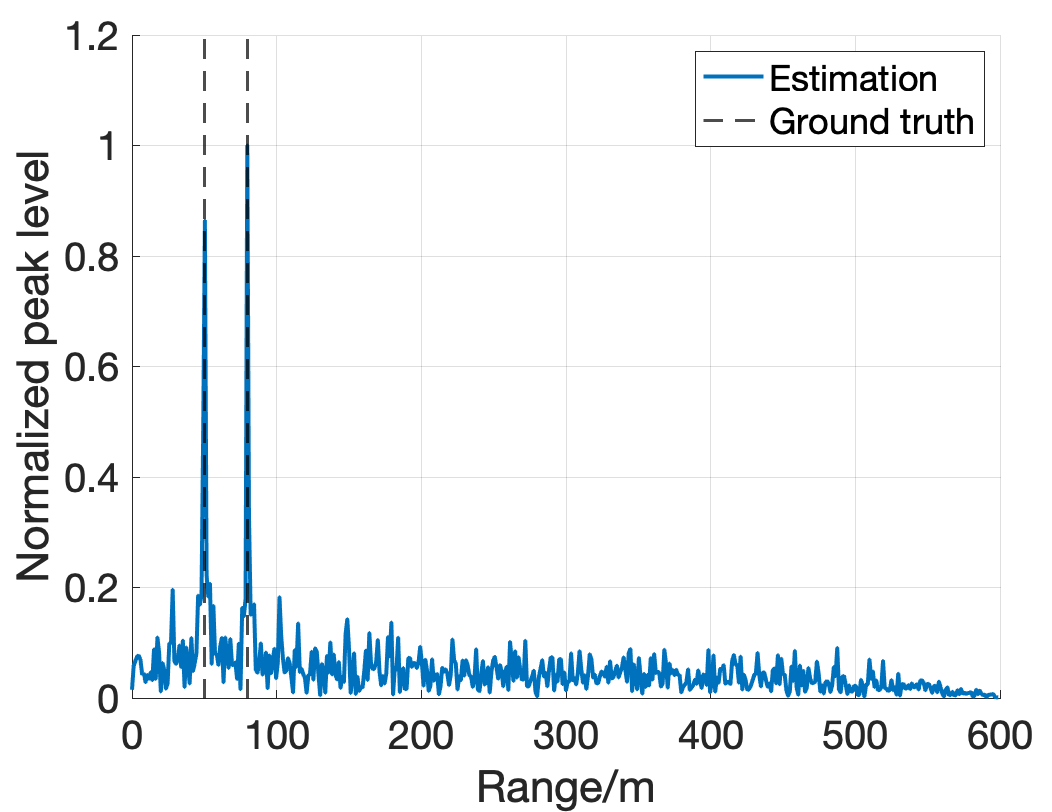}
    \caption{Range cross correlation result of two closely placed targets.}
    \label{fig:ranges}
    }
\end{figure}

\begin{figure}
\centering
\includegraphics[width = 7cm]{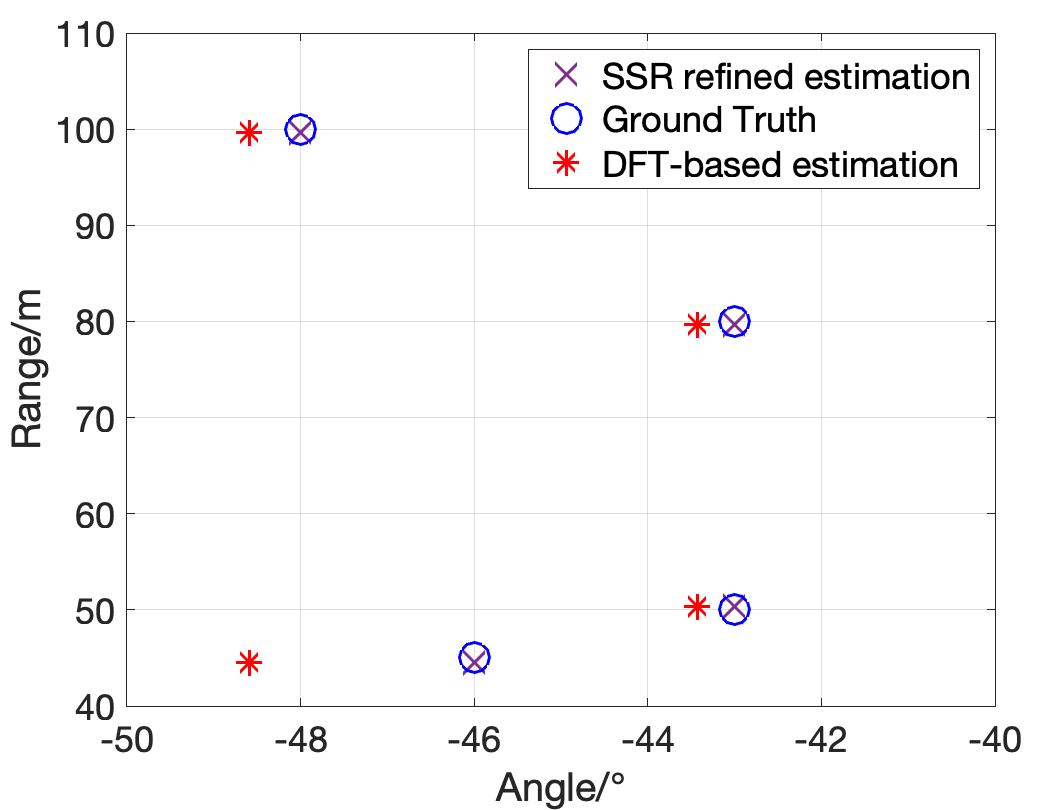}
\caption{Refined estimation results with  private subcarriers and virtual array.}
\label{fig:refined_est}
\vspace{-3mm}
\end{figure}

Following  Algorithm~\ref{alg:angle-range} of Section~\ref{sec:angle-range},  we first obtain coarse estimates of the target angles based on the physical array of $N_r=32$ antennas, using all subcarriers. 
Subsequently, we obtain range estimates within the estimated angle bins, via \eqref{Aprime}-\eqref{rangeind} based on the peaks of the coarse angle estimation (see Section \ref{sec:Range&Doppler}).
The results are marked in Fig.~\ref{fig:refined_est} by  red asterisks. As one can see that, due to the low angle resolution, only $2$ different angle bins are found, i.e., angles $-43.43\degree$ and $-48.59\degree$, with each bin containing $2$ targets. 
\textcolor{black}{
Recall that the coarse angle estimation is repeated on all subcarriers to exploit the frequency diversity. With a large number of subcarriers in an OFDM signal, one could find all the occupied angle bins.
However, if we only use one subcarrier for coarse angle estimation, we could miss one of the two occupied angle bins and thus miss the two targets within that angle bin. }

Fig.~\ref{fig:ranges} shows the cross-correlation result (see \eqref{rangeind}) in the angle bin corresponding to $-43.43\degree$. One can clearly see two narrow peaks at $50.39m$ and $79.69m$,  while the ground truth is $50$m and $80$m. 
Note that, since the ranges are all positive numbers, only the non-negative $x$-axis is shown in the figure. 
{In \eqref{eq: range_A}, both the private  and shared subcarriers are used in range estimation, thus allowing for  maximum range resolution. 

Indeed, 
the  obtained range estimates are correct, but the angle estimates are not matched correctly with the targets. The ground truth is also shown in the Fig.~\ref{fig:refined_est}.}
Using the $4$ range estimates we constructed and solved the  SSR problem  of \eqref{base_matrix}. $3$ angle bins were found,  i.e.,  $-43\degree$, $-46\degree$ and $-48\degree$.
The obtained angles were matched to the columns of the SSR basis matrix, and thus  to the ranges obtained so far. Based on that matching we found the angle-range pairs $(-43\degree,50.39m)$, ($-43\degree$,$79.69m$),   ($-46\degree$,$44.53m$) and ($-48\degree$,$99.61m$).
The results are shown in Fig.~\ref{fig:refined_est} marked by purple crosses. 
Re-estimating the ranges based on those angles does not change the results.
In this experiment, all targets' ranges are found in the initial estimation, thus only one more iteration was needed to refine the results (Steps $2$ and $3$ in Algorithm~\ref{alg:angle-range}), while
another iteration was used to verify the convergence of the iterative algorithm, i.e.,  no new targets were revealed. Thus, after a total of $2$ iterations of estimation, all targets were estimated with high resolution. 

The proposed angle-range alternating estimation algorithm is very useful in dealing with the coupling of parameters and works well even with a coarse estimate as initialization. As shown in the previous results, not only can the formulated virtual array  improve the angular resolution, but it can also help  match the refined angle estimates with estimated ranges.

\begin{figure}
\centering
\includegraphics[width = 7cm]{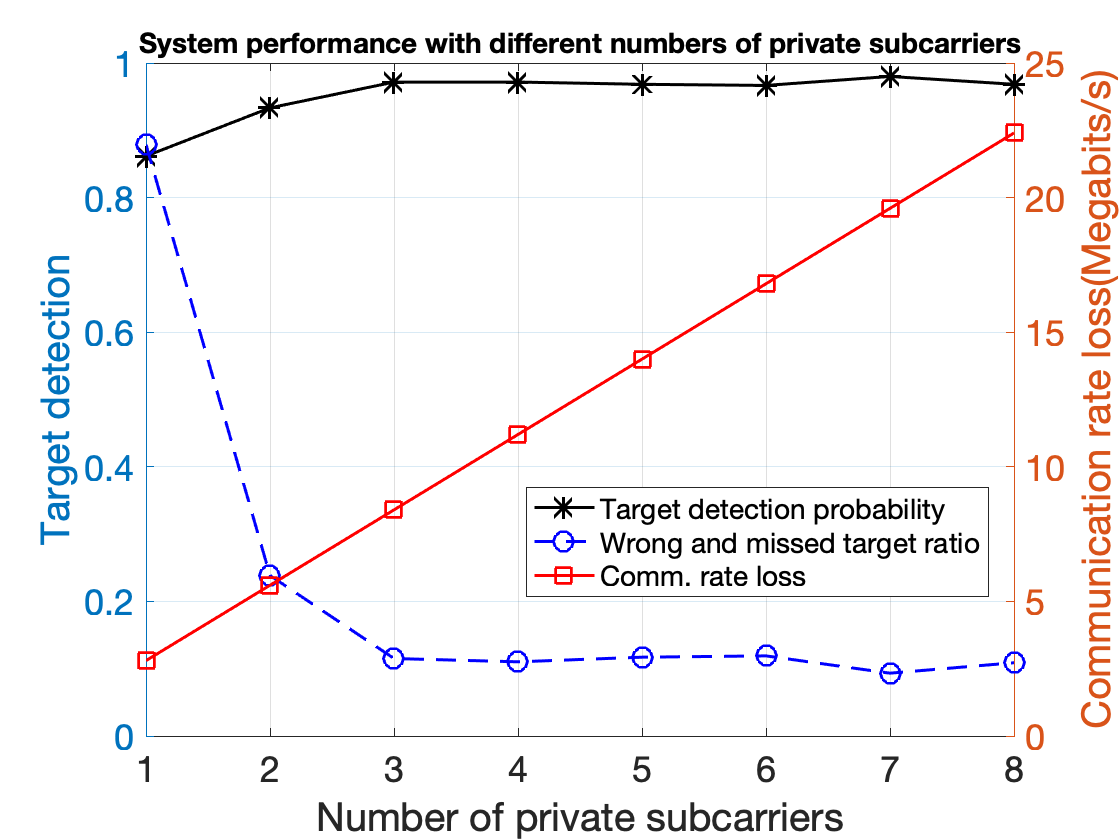}
\caption{System performance with different number of private subcarriers.}
\label{fig:private_compare}
\vspace{-3mm}
\end{figure}
\textcolor{black}{
 In order to evaluate the trade-off between radar performance and communication rate in \textcolor{black}{SS-DFRC}, Monte Carlo experiments were conducted with different number of private subcarriers. In each experiment, $6$ point targets were randomly generated and detected following the proposed Algorithm~\ref{alg:angle-range} with different number of private subcarriers. 
 The proposed iterative algorithm, it takes on average fewer than $1.5$ iterations to reach the convergence in all numbers of private subcarriers.
 For a detected target, if the error in any of the parameters exceeds the corresponding resolution, the estimation is classified as wrong.
The  probabilities of successful detection, where all targets  are detected and estimated correctly,  and the ratio between the number of 
wrong/missed  targets and total detected targets  are shown in Fig.~\ref{fig:private_compare}. 
 The  communication rate loss is also shown on the same figure. From the figure one can see that, despite the similar convergence speed, the target estimation can recover most of the targets when the number of private subcarriers is close to the number of transmitting antennas, with low probability of estimating wrong targets, or missing targets. However, the loss of communication rate could be as much as $2.8$ Megabits/s for changing one shared subcarrier into a private one under the current setup. From Fig.~\ref{fig:private_compare}, it can be seen that it is better to use $M=3$ private subcarriers for achieving a good radar performance and incurring small communication rate loss. Further, a small number of $M$ would lead to a shorter effective virtual array, thus reducing the size of SSR problem (see \eqref{virarr}) and the time required to solve it.}

\textcolor{black}{We should note that} the above angle and range estimation and refinement process required one OFDM symbol only, while the estimation of velocity requires multiple OFDM symbols. Based on the corresponding range peaks, we can estimate the velocities of targets via \eqref{veloind} and the results are $21.36m/s$,$9.16m/s$, $12.21m/s$ and $-9.16m/s$.

\begin{figure}
    \centering
    \includegraphics[width = 7.3cm]{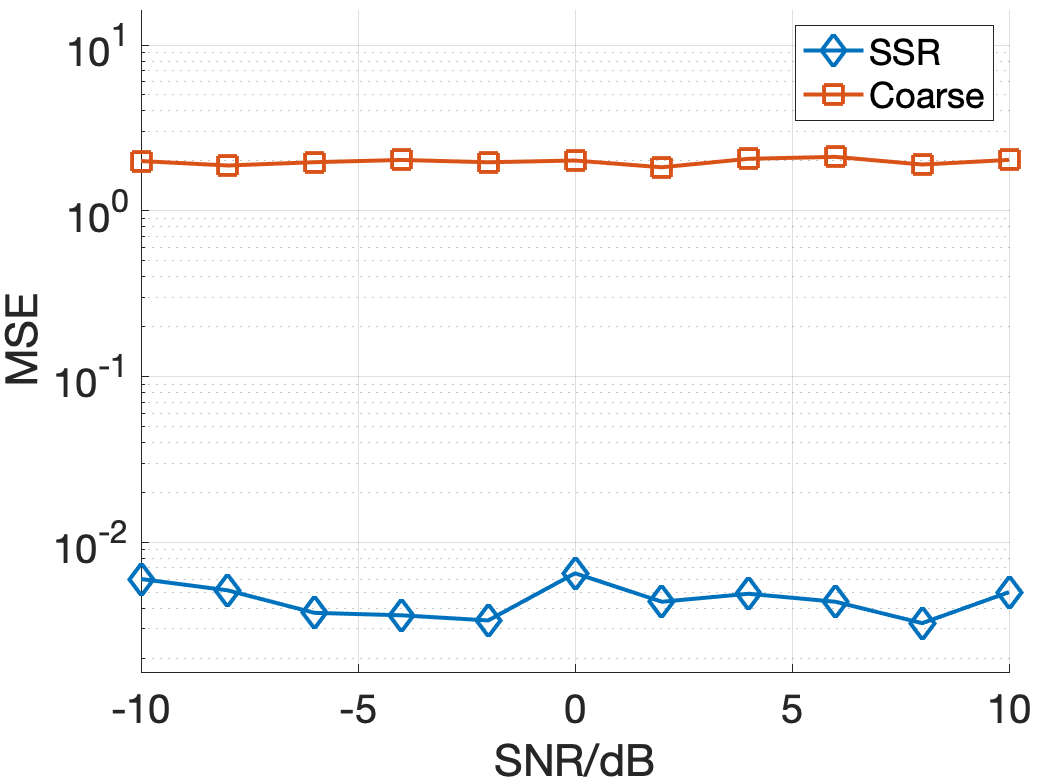}
    \caption{MSE on coarse angle estimation and SSR refined angle estimation}
    \label{fig:mse_angle}
    \vspace{-2mm}
\end{figure}

\begin{figure}
    \centering
    \includegraphics[width = 7cm]{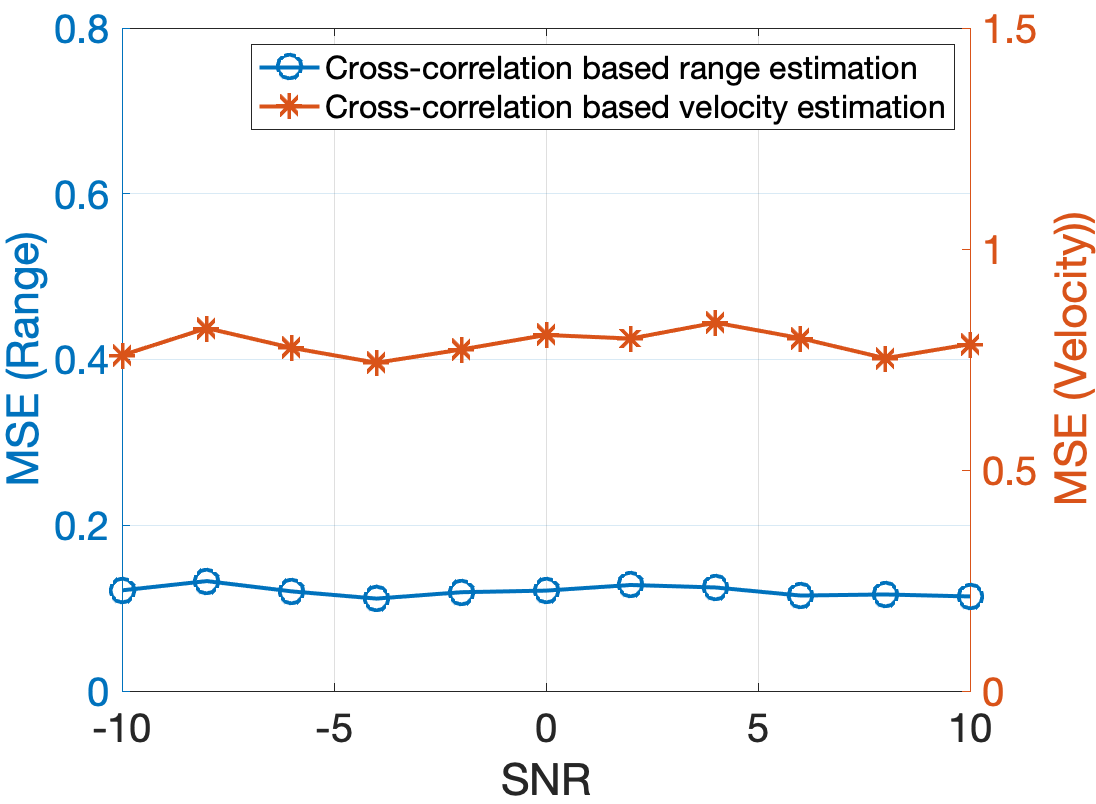}
    \caption{MSE on cross-correlation based range and Doppler estimation}
    \label{fig:mse_rD}
    \vspace{-3mm}
\end{figure}

\textcolor{black}{
In order to quantify the  improvement of angle estimation due to the use of private subcarriers, Monte Carlo experiments were conducted with $4$ private subcarriers under different SNR settings. $500$ experiments were repeated for each SNR. In each experiment, $1$ target is randomly  generated and detected. The performance of cross-correlation based range and Doppler estimation are also investigated. As shown in Fig.~\ref{fig:mse_angle}, the mean squared error (MSE) of SSR angle estimation is much smaller than that of coarse angle estimation based on the physical receive array. Although the use of private subcarriers is at a cost of communication rate, the improvement on sensing performance is significant. The MSE for cross-correlation based range and Doppler estimation is shown in Fig.~\ref{fig:mse_rD}.  
The range and Doppler estimation is based  on  finding  peaks in the cross-correlation result, while the noise only changes the sidelobe level. Thus, by properly setting up the peak detection threshold, the proposed method can be robust to different noise levels as shown in Fig.~\ref{fig:mse_rD}.}

 \subsection{Communication performance}

We consider a  communication receiver with $N_c = 64$ receive antennas. 
%
The channels are simulated following \eqref{eq:channel}, where the communication receiver is as distance $R_c = 50$ meters  from the radar transmitter. 
For the direct path between the transmitter and communication receiver, the departure angle is $30\degree$  and the incident angle is $-45\degree$.
The channel coefficients follow the same distribution as in the radar channel.  The departure and incidence angles of scatters are random, ranging from $-90\degree$ to $90\degree$ and \textcolor{black}{the corresponding coefficients, $c_k$, are random complex numbers with  mean equal to $0.1$ and variance equal to $1\times10^{-2}$.}

\begin{figure}
    \centering
    \includegraphics[width = 6.5cm]{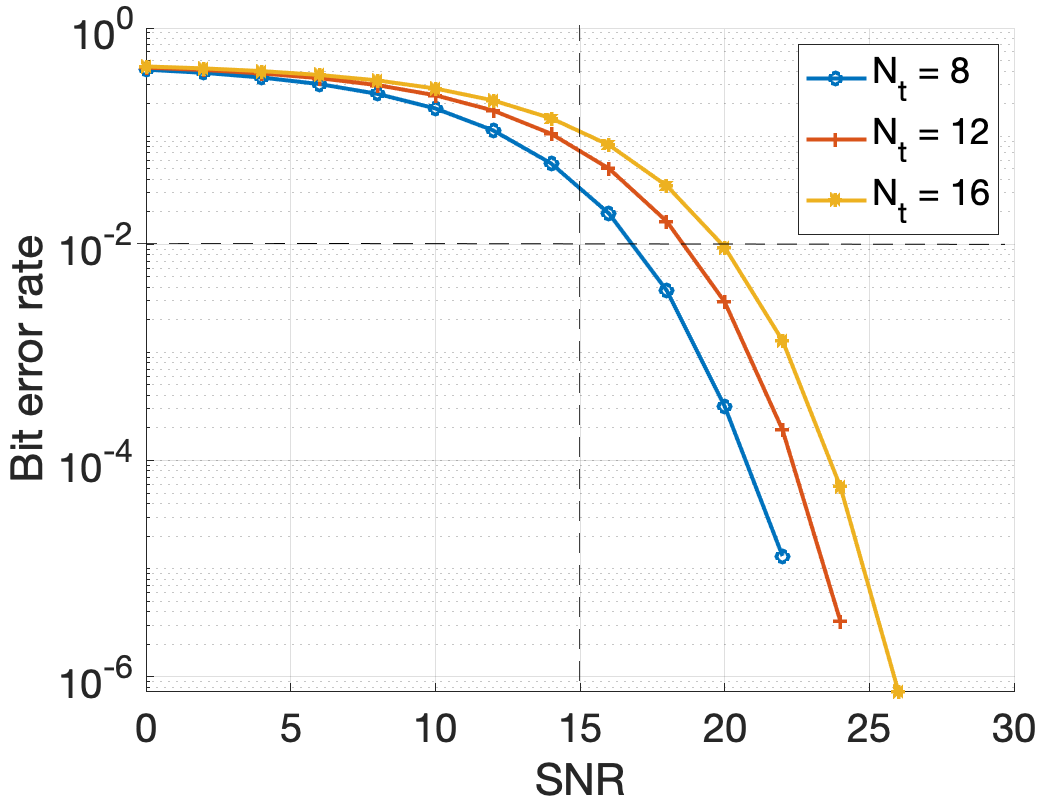}
    \caption{Bit error rate on the shared subcarriers for different number of transmit antennas.}
    \label{fig:ber}
\end{figure}

Assuming  that the  channel is known at the communication receiver, 
the data symbols can be recovered and mapped back to binary bits via  QPSK demodulation. 
The bit error rate (BER) of the proposed DFRC system   under different SNRs and with different number of transmit antennas is shown in Fig.~\ref{fig:ber}. 
One can see that fewer transmit antennas lead to  smaller BER. 

\begin{figure}
    \centering
    \includegraphics[width = 6.5cm]{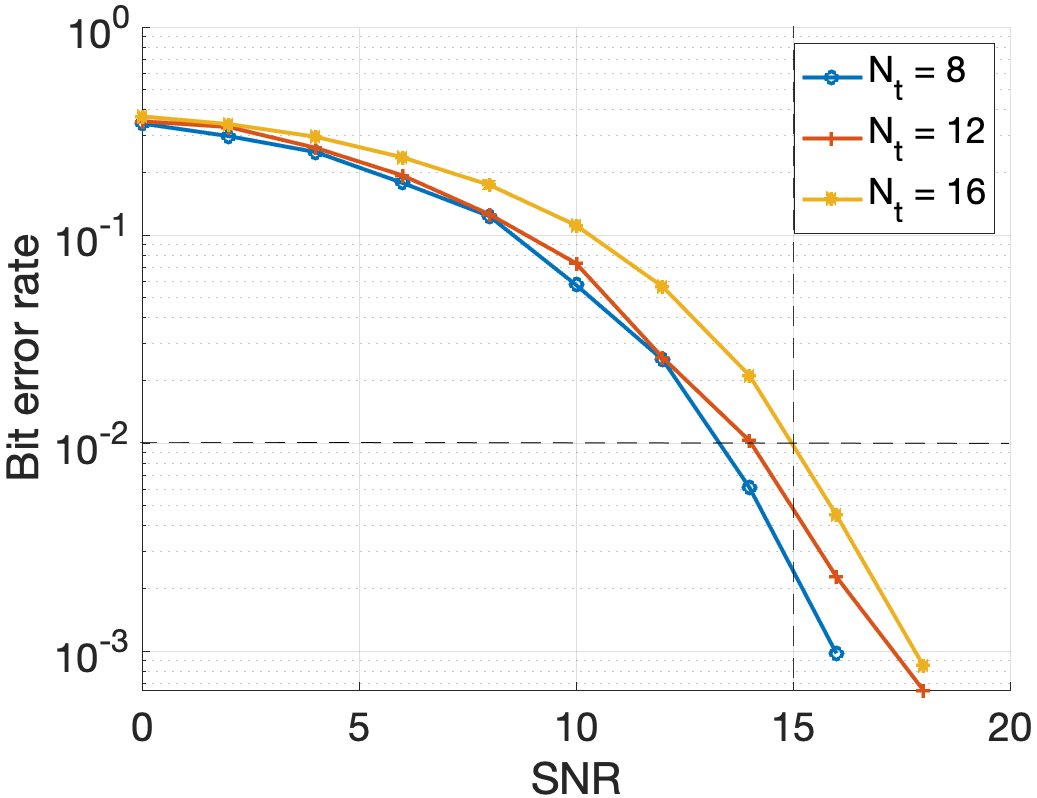}
    \caption{Bit error rate on the private subcarriers for different number of transmit antennas.}
    \vspace{-3mm}
    \label{fig:ber_ps}
\end{figure}

{
Here, we assume that the communication receiver knows the location of private subcarriers. 
}
Since there is no precoding is  used on the private subcarriers, the decoding amounts to finding the element in the constellation diagram, which after multiplication by the corresponding channel vector yields the smallest distance to the received symbol vector. The corresponding BER result is shown in Fig.~\ref{fig:ber_ps}. On the shared subcarriers, $N_t =8$ has the lowest BER, {while on the private subcarriers $N_t =8$ and $N_t =12$ have similar low BER when SNR is small. Compared to Fig.~\ref{fig:ber}, one can see that the BER on the private subcarriers is smaller than that on the shared ones.
}
%
Given that the communication rate increases with $N_t$, $N_t = 12$ is a better choice in terms of overall communication performance.

\begin{figure}
\centering
\includegraphics[width =6.5cm]{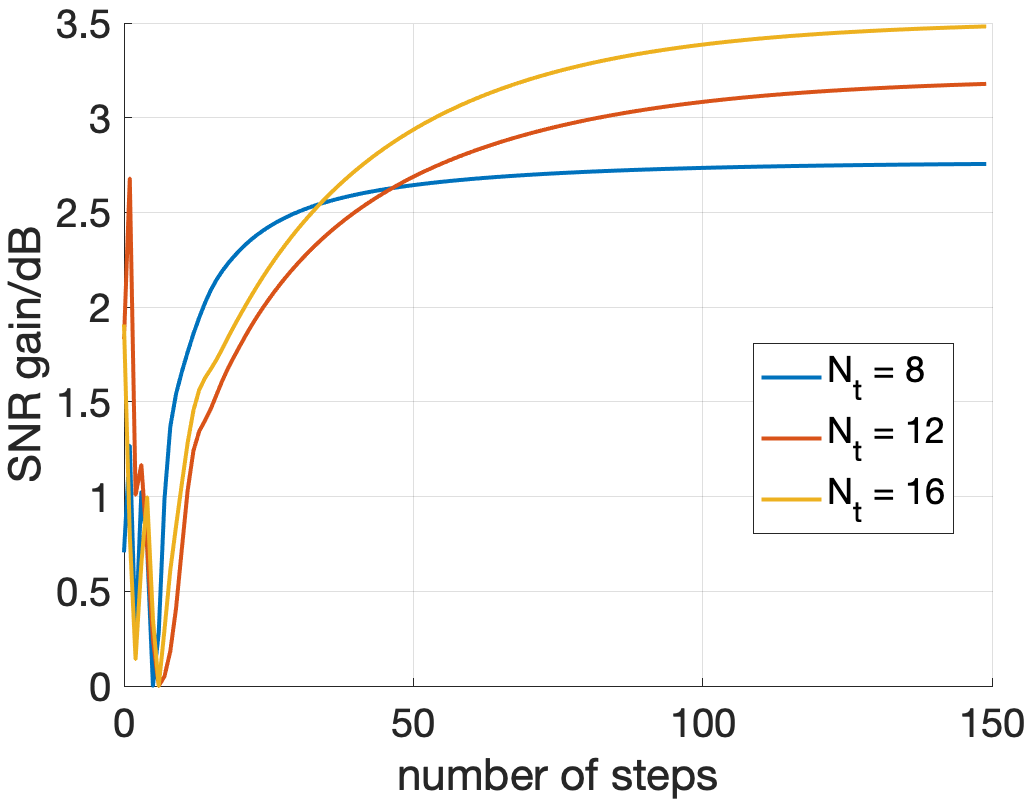}
\caption{SNR gain at the communication receiver  for different number of transmit antennas.}
\vspace{-3mm}
\label{fig:multiple_snr}
\end{figure}

The SNR gain for different number of transmit antennas is plotted in Fig.~\ref{fig:multiple_snr}. 
One can see that {more antennas result in higher SNR gain and also enable good beampattern performance. However, more antennas also result in  higher BER.}


Under the configuration provided in the table~\ref{table:parameters}, the maximum bit rate of the system  is $3.277$ Gigabits per second, when $N_t = 16$ antennas are used, while the loss of bit rate due to the use of  private subcarriers is $6$ Megabits per second.

\vspace{-2mm}
\section{Acknowledgments}
\textcolor{black}{
The authors would like to thank the associate editor and anonymous reviews for their insightful comments, and Prof. Joseph Tabrikian for useful discussions.}

\section{Conclusion}
We have proposed a wideband OFDM-MIMO DFRC system,
where subcarrier sharing is employed to increase the communication bit rate. 
Our formulation has taken into account a frequency selective fading channel model both for the sensing and communication environments. 
Regarding the sensing functionality, it has been shown that a coarse angle estimate can be first obtained  based on one snapshot of the receive array, and based on that, range information can be obtained within each occupied angle bin with maximum resolution via cross-correlation operations. 
We have also shown that the obtained angles  can be refined by exploiting an  effective virtual array, synthesized based on a set of private subcarriers. With the refined angles, the range estimation can be further improved,  which in return can further improve the angle estimation on private subcarriers. Based on that idea, we have proposed an iterative algorithm to improve both the angle and range estimation, which has been shown to work well,  successfully  pairing estimated ranges and angles. 
\textcolor{black}{In order to trade off the improvement on target estimation and the loss on communication rate, the number of private subcarriers has to be chosen carefully.}
After receiving several OFDM symbols, the Doppler frequencies can be estimated based on the range cross correlation peaks and thus pair with angle and range estimates.


{The communication data symbols can be estimated via a least-square method. As shown in the experiment results, increasing the number of transmitting antennas enables a higher SNR gain. However, the BER on shared subcarriers for a larger transmitting array is also higher, thus one needs to carefully choose the number of transmit antennas to ensure good overall performance. }



The   transmit precoding matrix have been {obtained via a gradient decent} approach, aiming to approximate a desired probing pattern and maximize the SNR at the communication receiver simultaneously. 

The functionality of our proposed DFRC system has been demonstrate via simulations. It has been shown that the proposed system has good sensing performance and  sustains high communication rate.




\ifCLASSOPTIONcaptionsoff
  \newpage
\fi



%


\bibliographystyle{IEEEtran}
\bibliography{ref}

%








\end{document}